\documentclass[aps,pre,onecolumn,superscriptaddress,amsmath,amssymb]{revtex4-2}

\usepackage[utf8]{inputenc}
\usepackage{amsmath,amsthm}
\usepackage{graphicx}
\usepackage[colorlinks = true,
            linkcolor = blue,
            urlcolor  = blue,
            citecolor = blue,
            anchorcolor = blue]{hyperref}
\usepackage{braket}
\usepackage{xcolor}
\usepackage{easyReview}
\newcommand{\bea}{\begin{eqnarray}}
\newcommand{\eea}{\end{eqnarray}}

\global\long\def\ga{\gamma} 
\global\long\def\De{\Delta} 
\global\long\def\th{\theta}
\global\long\def\th{\theta}

\global\long\def\ell#1{\theta_{#1}}

 \global\long\def\ka{\kappa}
\global\long\def\si{\sigma}
\global\long\def\vfi{\varphi}

\global\long\def\eps{\epsilon}
\global\long\def\al{\alpha}
\global\long\def\be{\beta}
\global\long\def\ga{\gamma}

\global\long\def\no{\nonumber}

\theoremstyle{thm@}

\theoremstyle{remark}

\def\pp{\mathbf{p}}
\def\qq{\mathbf{q}}

\newcommand{\eE}{\mathsf{e}}

\begin{document}

%\title{Spin helices decay under the $XXZ$ Heisenberg spin-$\frac12$ chain dynamics: Theory}
\title{Universality in  relaxation of spin helices under the $XXZ$-  spin chain dynamics}
%\title{Time evolution of spin-helix states  under the $XXZ$  spin-$\frac12$ chain dynamics}
%$XX0$  spin-$\frac12$ chain }
\author{Vladislav Popkov}
\affiliation{Faculty of Mathematics and Physics, University of Ljubljana, Jadranska 19, SI-1000 Ljubljana, Slovenia}
 \affiliation{Department of Physics,
  University of Wuppertal, Gaussstra\ss e 20, 42119 Wuppertal,
  Germany}
\author{Marko \v{Z}nidari\v{c}}
\affiliation{Faculty of Mathematics and Physics, University of Ljubljana, Jadranska 19, SI-1000 Ljubljana, Slovenia}
\author{Xin  Zhang}
\affiliation{Beijing National Laboratory for Condensed Matter Physics, Institute of Physics, Chinese Academy of Sciences, Beijing 100190, China}

%\author{Vladislav Popkov}
%\affiliation{Faculty of Mathematics and Physics, University of Ljubljana, Jadranska 19, SI-1000 Ljubljana, Slovenia}
% \affiliation{Department of Physics,
  %University of Wuppertal, Gaussstra\ss e 20, 42119 Wuppertal,
  %Germany}

\begin{abstract}
We describe dynamics of transverse spin-helix state (SHS) -- a product state with spatially rotating magnetization -- under anisotropic Heisenberg spin chain evolution. Due to experimental relevance we especially focus on magnetization dynamics. At long times the $U(1)$ symmetry of the Hamiltonian is restored, leading to the decay of transverse magnetization, which can be described as an exponential decay of a spatially harmonic profile. We show that the dependence of the short  and intermediate-time decay timescale, which in principle depends on all different parameters, like the wavevector of the initial helix, the anisotropy, etc., can be described well by a single scaling function. We also briefly discuss the evolution of magnetization current.  
\end{abstract}
\maketitle

New experimental techniques allow to create novel quantum states  with unusual properties.  Among them,   helices in Heisenberg magnets with a
uniaxial anisotropy were produced and manipulated in cold atom experiments  \cite{2014-SHS-Experimental,SHS-Ketterle,2020NatureSpinHelix,2021KetterleTransverse}.   
Simplicity of creation and chiral properties of the quantum helices,  their nontrivial topology and large magnetization current makes quantum helices attractive 
for potential applications in spintronics and quantum computing.  With the aid of $2D$ helices one finds nonequilibrium universality features \cite{2022-FarFromEquilibriumEniversality-2D}.
Using helicity degrees of freedom as qubits was recently 
discussed in \cite{2021SkyrmionsQubits}. 
Stability of $1D$ helix states to external noise can exceed the  stability of the ground states as was argued in \cite{2023Posske}. 
The helices can  be prepared from most simple  initial ``vacuum"-like  states by application of resonant fields  \cite{2022-DM-SHS} or by an adiabatic rotation \cite{2019-PosskeWinding}.
Due to simple structure,  helices can be also  maintained by dissipation,  namely they become dark states under properly chosen
local dissipative protocol,  affecting only boundary spins \cite{2016SHS-DissCarlo,2017SHS-DissGunter}.  Finally,  $1D$ helices  have a nontrivial content in terms of 
quasiparticles: in the Bethe Ansatz framework,  helices are formed by exotic quasiparticles carrying zero energy and finite momentum,  the so-called
phantom Bethe  excitations \cite{SHS-Phantom}.

Our purpose  is to set up a general theoretical framework for a problem,   addressed in a beautiful experiment \cite{SHS-Ketterle,2021KetterleTransverse}.
The problem is to describe the time evolution of helices with arbitrary wavevector  under a $XXZ$ spin $\frac12$ coherent dynamics
characterized by $z$-axis anisotropy $\De$.  Note that unlike the helices with modulation in $XZ$-plane, discussed in \cite{2020NatureSpinHelix,2022XZ-SHS,SHS-Hydro} we treat 
the transverse helices with modulation in $XY$-plane: these two helix types behave completely differently under the $XXZ$ dynamics; in particular,  transverse helices can be long-lived quantum states \cite{SHS-Ketterle,2022-QuantumScars-Review}.
Central object of interest in experimental studies  \cite{SHS-Ketterle,2021KetterleTransverse}   is a decay rate of  transverse 
helix amplitude which is calculated from raw data using some ad-hoc or phenomenological fit function.  Here 
we show that the problem exhibits  scaling features leading to universal behaviour of the decay rate. 
More specifically, we show that the rate of change of the transverse amplitude
 $\ga(Q,\De)$ of a helix with wavevector $Q$ under the $XXZ$ dynamics with anisotropy $\De$ exhibits self-similar scaling 
\begin{align}
\frac{\ga(Q,\De)}{\cos Q} =  \ga\left( 0,\frac{\De}{\cos Q} \right), \label{eq:Scaling0}  
\end{align}
valid for  short time and intermediate time window. Range of validity of (\ref{eq:Scaling0}) depends on system parameters. For $\De=0$
(non-interacting fermions)   the scaling (\ref{eq:Scaling0})  is exact, and moreover, we find multipoint correlations to satisfy their own scaling relations. For other regimes, we supply arguments that (\ref{eq:Scaling0}) holds at least up to times where the transverse SHS amplitude drops by a factor of $2$, which is the most experimentally relevant time window.  Note that the decay rate $\ga$ always depends on time via its definition.  Here we accept a definition (\ref{def:ga}) based on a threshold,  in order to make direct comparison with experiment \cite{SHS-Ketterle}.

%For the moment, such understanding is lacking.
In the following,  we set up the problem and derive  conceptually important properties and symmetries.  We find that two real functions 
(the amplitude and the phase ) fully describe the temporal dynamics of one-point correlations,  and investigate these functions numerically and analytically.  In the free fermion case we find remarkable scaling properties for all equal time observables.  Then  we treat general 
$XXZ$ case,  and derive the  scaling relation (\ref{eq:Scaling0}) leading to data collapse of experimentally accessible quantity, a half-amplitude decay rate.   We compare our findings  with existing experimental data. 
%\rev{We find self-similar scaling behaviour in far from equilibrium quantity, the decay rate of the helix amplitude for 
%different wavelengths }
 At the end we discuss the evolution of the magnetization current and give verifyable quantitative predictions.

\section{Setup of the problem}
We are interested in  the temporal evolution of a $1D$  spin helix state (SHS)

\begin{align}
&\ket{\Psi_{Q,\th,\vfi}}=\bigotimes_{n=0}^{N-1} 
\binom {e^{-i\frac{Qn+\vfi}{2} }\cos \frac{\th}{2}}{e^{i\frac{Qn+\vfi}{2}} \sin\frac{\th}{2}} \label{eq:SHS} \\
&Q N = 0 \ \mod \ 2\pi, \label{eq:commensurate}
\end{align}
describing a helix spiral with period $2\pi/Q$ in lattice units, constant polar angle $0\leq \th \leq\pi$,
and phase $\vfi$.  
In the following we shall also use a shorthand notation $\ket{\Psi_{Q}}\equiv \ket{\Psi_{Q,\th,\vfi}}$
and especially,  for  a spatially homogeneous version of SHS ($Q=0$), we shall use  $\ket{\Psi_{0}}$,
\begin{align}
&\ket{\Psi_{0}} =\binom {e^{-i \frac{\vfi}{2}} \cos \frac{\th}{2}}{e^{i \frac{\vfi}{2}} \sin\frac{\th}{2}}^{\otimes_N}.\nonumber
\end{align}
Chiral and homogeneous  SHS   are related $\ket{\Psi_{Q}} = U_Q \ket{\Psi_{0}}$ via 
simple transform $U_Q$ given 
in (\ref{f'}).
The SHS (\ref{eq:SHS}) evolves via coherent quantum dynamics described by
the  $XXZ$  Hamiltonian with periodic boundary conditions
\begin{align}
&H=\sum_{n=0}^{N-1} h_{n,n+1}, \quad N+n\equiv n,\label{Ham} \\
&{h}_{n,n+1}={\sigma}_n^x\sigma_{n+1}^x+\sigma_n^y\sigma_{n+1}^y+ \De\ (\sigma_n^z\sigma_{n+1}^z-I).\nonumber
\end{align}
with $z-$ axis exchange anisotropy $\De$.  Eq.(\ref{eq:commensurate}) provides commensurability of a SHS in a periodic system.
The time evolution $e^{-iHt}\ket{\Psi_{Q,\th,\vfi}}$
is characterized by  expectation values for  observables,  denoted as
\begin{align}
&\langle A(H,t)  \rangle_{Q} \equiv  \bra{\Psi_{Q,\th,\vfi}} e^{iHt}  A\, e^{-iHt}  \ket{\Psi_{Q,\th,\vfi}},
\label{eq:observable}
\end{align}
(explicit dependence on $\th,\vfi$ is omitted at the LHS of  (\ref{eq:observable}) for brevity), where $A$ is the operator of an observable. 
Further on, we also omit unnecessary variables in $\langle A(H,t)\rangle_{Q}$ and $\ket{\Psi_{Q,\th,\vfi}}$  whenever it will not lead to misunderstanding (omitted variable is the  same for 
all  terms in an equality).

Particularity of the state (\ref{eq:SHS}) is its chirality,  characterized by  integer $QN/(2\pi)$,  the winding number in the clockwise direction, and 
current of $z-$magnetization  
\begin{align}
\langle j^z(t=0) \rangle =2\sin^2 \th \sin Q,\quad  j^z=2(\si_n^x \si_{n+1}^{y}-\si_n^y \si_{n+1}^{x}) .
\label{eq:jSHS}
\end{align}
In addition,  for
$\cos Q=\De$,  SHS  (\ref{eq:SHS}) is an eigenvector of $H|_{\De=\cos Q}$ with eigenvalue $0$ \cite{SHS-Phantom,Phantom-Long}, 
so the time evolution 
(\ref{eq:observable}) can be viewed as a result of a quench  from $H|_{\De=\cos Q}$ to $H$ with arbitrary anisotropy at time $t=0$. Finally,
SHS can be prepared in  experiments by manipulating  homogeneously polarized  
 equidistantly  separated qubits with a magnetic field gradient \cite{SHS-Ketterle}.

%\section{SHS decay: time asymptotic form in the thermodynamic limit $N\rightarrow \infty$}
%\label{sec:SHSdecay}

Under standard  assumptions, see Appendix \ref{App:SHSdecay} for details, one can assume that a  a $U(1)$- invariant operator like $H$ on an infinite lattice
will impose $U(1)$ symmetry on any   its subsystem of finite size  asymptotically in time, leading,  specifically,  to 
decay of transversal magnetization: 
\begin{align}
&\lim_{t\rightarrow \infty} \lim_{N\rightarrow \infty} \langle \si_n^\pm (t)  \rangle =0, \quad \forall n \label{eq:corr1}
\end{align}
where $\si_n^\pm = \frac12 (\si_n^x \pm i \si_n^y)$.

\section{General properties of SHS observables under $XXZ$ evolution}
\textit{Property I.  Relation  between spatially shifted observables.}

Let $A_{n}$   and  $A_{n+1}$ be the same operator,  shifted by one site (the operator itself can  act on arbitrary number of sites).  
Then, 
\begin{align}
&\langle A_{n+1}\rangle= \langle V_Q^\dagger  A_{n} V_Q\rangle, \label{eq:shift}\\
&V_Q = \bigotimes_{n}   e^{-i  \frac{Q}{2} \si_n^z}. \nonumber
\end{align}
For a proof,  denote by $T$ an operator of a shift by one lattice site to the right. Obviously,  $[T,H]= [V_Q,H] =0$.  %In the following,  we shall use  shorthand notation $\ket{\Psi}$ for $\ket{\Psi_{Q,\th,\vfi}}$,  and $H$ for $H_\De$ whenever it will not lead to misunderstanding.
Using easily verifiable relation
\begin{align*}
&T \ket{\Psi_Q} =  V_Q \ket{\Psi_Q},
\end{align*}
we obtain 
\begin{align*}
& \langle A_{n+1}(t)\rangle = \bra{\Psi_Q} T^\dagger e^{i H t}  \  A_n \  e^{-i H t} T \ket{\Psi_Q} \\
& = \bra{\Psi_Q}   V_Q^\dagger  e^{i H t}  \  A_n\  e^{-i H t}  V_Q \ket{\Psi_Q}\\
&= \bra{\Psi_Q}    e^{i H t}  V_Q^\dagger  \  A_n  \ V_Q  \  e^{-i H t}  \ket{\Psi_Q}, 
\end{align*}
i.e.  (\ref{eq:shift}). 
Iterating (\ref{eq:shift}) $k$ times we get $\langle A_{n+k}\rangle = \langle (V_Q^{-k} A_{n} \ V_Q^k)\rangle$ for an expectation value of 
an operator shifted by $k$ lattice units.  

%Another useful property relates observables of time-evolved SHS with arbitrary wavevector $Q$ to those
%computed for time-evolved homogeneous state $Q=0$:

\textit{Property II-- scaling relation.   Relation  between expectations calculated with  homogeneous state $\ket{\Psi_0}$ and with chiral state $\ket{\Psi_Q}$.}

Let $A$ be an arbitrary operator   and let
  $Q$ satisfy (\ref{eq:commensurate}). Then,
\begin{align}
&\langle A(H,t) \rangle_Q = \langle A'(H',t) \rangle_0,\label{eq:Theorem}
\end{align}
where 
\begin{align}
& A' = U_Q^\dagger A \, U_Q, \quad U_Q= e^{-i \frac{Q}{2} \sum_{n=0}^{N-1} n \, \si_n^z}, \label{f'}\\
%&H' = \widetilde{H} \, \cos Q - \frac{\sin Q}{2} J,\quad \widetilde H=H|_{\De\to\De/\cos Q},\label{eq:HH'}\\
&H' = \cos Q \sum_{n=1}^N \left(\si_n^x \si_{n+1}^x + \si_n^y \si_{n+1}^y + \frac{\De}{\cos Q} ( \si_n^z \si_{n+1}^z -I) \right) +
\frac{\sin Q}{2} J,\label{eq:H'}\\
&J=2 \sum_n ({\sigma}_n^x\sigma_{n+1}^y-\sigma_n^y\sigma_{n+1}^x). \label{DM}
\end{align}
For a proof,  note that $\ket{\Psi_{Q}} = U_Q \ket{\Psi_{0}}$.
Inserting unity $ U_Q^\dagger U_Q=I$  in  proper places,  we obtain 
\begin{align}
& \langle A(H,t)\rangle_Q=\bra{\Psi_{0} }
e^{iH't} A' e^{-iH't} \ket{\Psi_{0}}= \langle A'(H',t) \rangle_0, \quad \mbox{where $X' =  U_Q^\dagger X U_Q$.} \nonumber
\end{align}
Each term $\si_n^\al \si_{n+1}^\al$ in $H$ becomes   $(\si_n^\al)'( \si_{n+1}^\al)'$ after the $U_Q$ transformation.  To simplify further,  we use 
 easily verifiable relations 
\begin{align}
& (\si_n^x)'=\cos(nQ)\ \si_n^x-\sin(nQ ) \ \si_n^y,\\
& (\si_n^y)'=\cos(nQ)\ \si_n^y+\sin(nQ ) \ \si_n^x,\\
&(\si_n^z)'=\si_n^z.
\end{align} 
Inserting the above into $H'$ and using trigonometric identities we get (\ref{eq:H'}). 

In the thermodynamic limit $N \rightarrow \infty$,  $Q$ is arbitrary.  Then,  
 (\ref{eq:Theorem}) allows to reduce a  problem of SHS evolution 
with arbitrary wavelength $Q$ to  evolution of a homogeneous 
state $Q=0$,  under the transformed Hamiltonian $H'$,  containing an additional  Dzyaloshinskii-Moriya  term  \cite{DM1,DM2},
proportional to the total current of magnetization $J$ (\ref{DM}).

In the following,  we apply 
the Properties I,II to study  one-point observables,  i.e.   spin helix magnetization profile,  and selected two-point observable,   the 
current of magnetization. 

\section{Decay of trasversal SHS components}

\begin{figure}[tbp]
\centerline{
\includegraphics[width=0.475\textwidth]{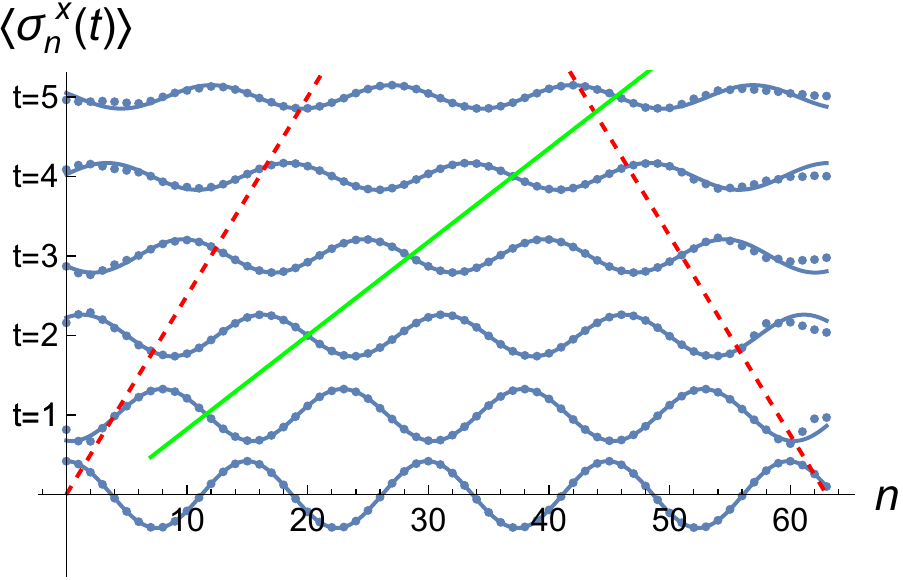}
}
\caption{
Decaying travelling wave:  $x$-component  of the magnetization profile at equal time intervals versus site number,  for
  $\th= \pi/5, Q=0.42, \De=0$.   Magnetization profiles at different times $t=0,1,\ldots ,5$ are shown 
with additional offsets for better visibility. 
Points are TEBD numerics for a chain with $64$ sites and open boundary conditions, while
 interpolating harmonic curves are given by Eq. (\ref{eq:SigmaX}).  
Open bondary conditions generate perturbations at the edges,  propagating in the bulk  with the velocity $v_{prop}\approx 4$ per unit time (red dashed lines). 
Green line shows the location of the constant phase,  and is given by $n=v t + a$ where $v \approx 8.5$ per unit  time. 
}
\label{Fig-TravellingWave}
\end{figure}

The  magnetization profile of SHS (\ref{eq:SHS}) at $t=0$ is harmonic in space, 
\begin{align}
&\frac{\langle \si_n^x(0) \rangle}{\sin \th} =  \cos(Q n+\vfi), \label{eq:X}\\
&\frac{\langle \si_n^y(0) \rangle}{\sin \th} =  \sin(Q n+\vfi), \label{eq:Y}
\end{align}
where $\th$ is the polar angle and $\vfi $ is the overall phase shift.  Applying  (\ref{eq:shift}) with  $A_n =\si_n^{\al} $
 and
 using  $e^{i \frac{Q}{2} \si^z}  \  \si^{\pm} e^{-i \frac{Q}{2} \si^z} = e^{\pm i Q}  \si^{\pm}$,  and    $e^{i \frac{Q}{2} \si^z} \ \si^{z} e^{-i \frac{Q}{2} \si^z}= \si^{z}$,  we obtain
\begin{align}
&\langle \si_{n+1}^\pm(t) \rangle= e^{\pm i Q } \langle \si_n^\pm(t) \rangle,\label{eq:SHSperiodicityArbTime}\\
&\langle \si_{n+1}^z(t) \rangle= \langle \si_{n}^z(t) \rangle.\label{eq:SHSperiodicityArbTime-Z}
\end{align} 
Eq. (\ref{eq:SHSperiodicityArbTime}) entails that the magnetization profile of SHS stays strictly harmonic in space at all times $t$,
see Fig.~\ref{Fig-TravellingWave} for an illustration,   
and can therefore be described via a rescaled amplitude $S_N(t)\equiv S_N(Q,\th,\De,t)$ and a phase shift $\phi(t)\equiv\phi(Q,\th,\De,t)$, as
\begin{align}
&\frac{\langle \si_n^x(t) \rangle}{\sin \th}  =  S_N(t)  \cos(Q n+\vfi -\phi(t)   ), \label{eq:SigmaX}\\
&\frac{\langle \si_n^y(t) \rangle}{\sin \th}  =  S_N(t)  \sin(Q n+\vfi -\phi(t) ), \label{eq:SigmaY}\\
&\langle \si_{n}^z(t) \rangle= \cos \th, \label{eq:SigmaZ}\\
&S_N(0)=1,\quad \phi(0)=0.\nonumber
\end{align}
Here,  Eq. (\ref{eq:SigmaZ}) follows from (\ref{eq:SHSperiodicityArbTime-Z}) and the fact that the $XXZ$ dynamics conserves total $z$- magnetization.

Both the $S_N$ and $\phi$ depend not only on time and system size $N$ but also on  $Q, \th,\De$,  while
there is no dependence on the overall phase $\vfi$ for obvious physical reasons. $S_N(t)$ and $\phi(t)$ satisfy relations
 \begin{align}
&S_N(t)= S_N(-t), \quad \phi(t) = - \phi(-t), \label{eq:parityS}\\
&S_N(Q,\th) = S_N(-Q,  \th) = S_N(Q, \pi- \th), \nonumber \\
&S_N(Q,  \De) =  S_N(\pi-Q, -\De),\,\,\mbox{for even $N$},\nonumber\\
&\phi(\pi-\th) = -\phi(\th) \label{eq:vThetaDep} \\
&\phi(Q) = \phi(-Q), \nonumber
\end{align} 
(omitted parameters are the same on both sides), imposed by symmetries of the Hamiltonian and the SHS,  see  Appendix \ref{App:PhaseShift}. 
In particular it follows from (\ref{eq:vThetaDep}) that 
\begin{align}
&\left. \phi(\th,t) \right|_{\th=\pi/2} = 0, \label{eq:vThetaPi2}
\end{align} 
i.e.  decay of the fully trasversal SHS  (the SHS with polarization lying in the $XY$-plane) is described  by just one  function $S_N(t)$
in (\ref{eq:SigmaX}),  (\ref{eq:SigmaY}). For $\th\neq \pi/ 2$, and $|\De|<1$, $\phi(t)$ quickly converges to $\phi(t)=v t$, which
allows to view the magnetization profile as a travelling wave, see 
green line in Fig.~\ref{Fig-TravellingWave}. Further details about the phase are given in sec.~\ref{sec:PhaseVelocity}.
Note that the amplitude $S_N(t)$ can be easily measured experimentally while  the phase is usually unknown.

We obtained explicit analytic form of $S_N(t)$ in several  cases: for $|\De| \rightarrow \infty $, see (\ref{S(t)DeLarge}), for $\th \rightarrow 0$ (see Appendix \ref{Theta0}), and for free fermion case  $\De=0$, via an explicit determinantal representation  (\ref{eq:S(t)-2}).

Further,  we are interested in thermodynamic limit $S_N(t)|_{N \rightarrow \infty} \rightarrow S(t) $,  since it is an experimentally
measurable quantity \cite{SHS-Ketterle}.  From Eq. (\ref{eq:corr1}) we expect the asymptotic decay of $S(t) \rightarrow 0$ at large times for any choice of  parameters,  apart from the case when the 
SHS is an eigenstate of $H$, i.e.  for $\cos Q = \De$.  We can determine 
early time behaviour of $S(t)$ or any other observable  via exact Taylor expansion, see (\ref{eq:TaylorSHS}), (\ref{eq:SigmaxTaylor}).

To obtain the observables at intermediate times $t=  O(1)$ we use  TEBD calculations, 
see Appendix ~\ref{App:DMRG}. 
The raw TEBD data are illustrated in Fig.~\ref{Fig-TravellingWave}.  We use the central area near the middle site $n=N/2$ for all measurements,
in order to avoid an influence of the borders.  The data thus obtained  effectively coincide 
with those from an infinite system.  
%Typical behaviour of $S(t)$ is shown in  Figs.~\ref{Fig-S(t)}-\ref{Fig-S(t)-De}.

The quality of the TEBD data for the bulk  can be checked  by monitoring deviations for bulk integrals of motion. For instance, 
$\langle\si_n^z \rangle $ component of the magnetization in the bulk must stay constant in space and time, see (\ref{eq:SigmaZ}).
Thus, we trust the DMRG data for $S(t)$ up to the times when  $\langle\si_n^z(t) \rangle-\langle\si_n^z(0) \rangle$ deviations in the bulk  start to appear in the middle of the chain, where we do the measurements. For  the  bond size $\chi=20$ this leads to parameter-dependent $t_{max}$, the typical value being $t_{max} \approx 3$, 
which is enough for our purposes,
 see Appendix~\ref{App:DMRG} for more details.  In addition, we checked the TEBD correctness directly by comparison with the exact result (\ref{eq:S(t)-2}).

The case $\De=0$ is special and deserves separate discussion.

\begin{figure}[tbp]
\centerline{
\includegraphics[width=0.5\textwidth]{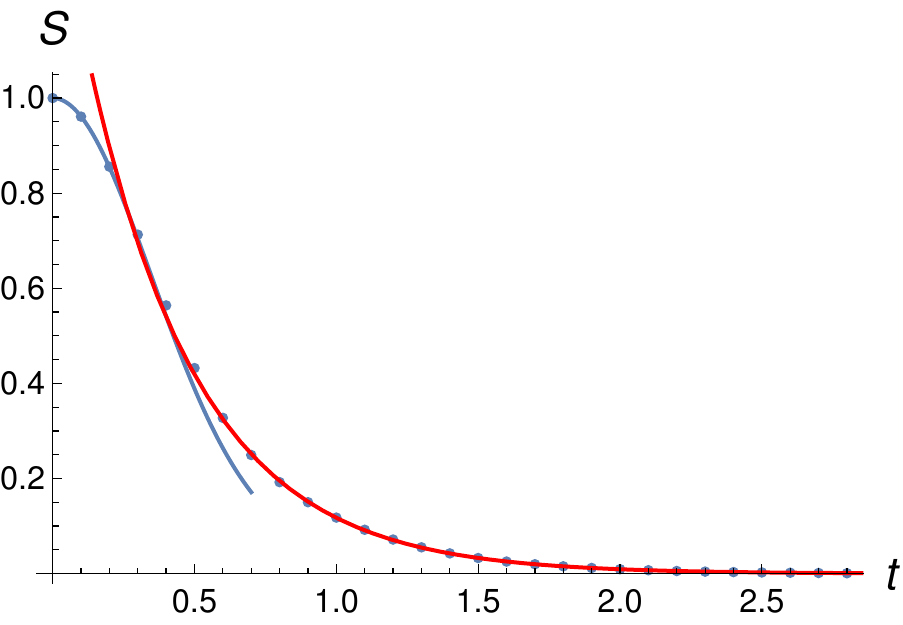}
}
\caption{
Decay of the amplitude $S(t)$ for homogeneous initial state $\ket{\Psi_0}$, with $\th=\pi/2$ under $XX$ dynamics ($\De=0$).  Points stem from  exact formula (\ref{eq:S(t)-2})
for $N=22$ sites.
Interpolating  red curve is given by $1.49\  \eE^{-2.54 t}$.  Interpolating blue curve is given 
by renormalized symmetric $\al-$ stable L\'{e}vy  distribution \cite{durrett2019probability} with $\al=1.8$.
}
\label{Fig-S(t)}
\end{figure}

%The next section deals with the free-fermion case that has peculiarities and is treated separately. 
\section{$XX$ case.  Scaling form of correlations for different SHS wavelengths} \label{Section;XX}

For free fermion case $H\equiv H_{XX} = \sum_{n} ({\sigma}_n^x\sigma_{n+1}^x+\sigma_n^y\sigma_{n+1}^y)$,
the total magnetization current $J$ from (\ref{DM}) is a constant of motion:
$[H_{XX},J]=0$.  
This renders the Dzyaloshinskii-Moriya term containing $J$ in (\ref{eq:H'}) irrelevant, if the initial state is an eigenstate of $J$. 
 Indeed, in our case one can  obtain $J \ket{\Psi_{0}} =0$ (no gradient, no current),  resulting in $e^{-iH_{XX}'t} \ket{\Psi_{0}}= e^{-iH_{XX}t \cos Q }\ket{\Psi_{0}}$. 
Consequently,  for $H\equiv H_{XX} $ the scaling property (\ref{eq:Theorem}) simplifies as
\begin{align}
& \langle A(H_{XX},t)\rangle_Q  = \langle A'(H_{XX},t\cos Q) \rangle_0. \label{eq:HxxTheorem}
\end{align}
An immediate consequence of (\ref{eq:HxxTheorem}) is a scaling relation
for the  SHS amplitude 
\begin{align}
&S_N(Q,t)=S_N(0,t\cos Q ),\label{eq:XX-Similarity}
\end{align}
i.e.  the curves $S_N(t)$ for different $Q$ differ just by rescaling of time! Note that (\ref{eq:XX-Similarity}) is
valid for any finite $N$,  provided commensurability of $Q$ (\ref{eq:commensurate}).  

For  even $N$ and $\De=0$, one  finds \cite{2023ChiralBasis} explicit  expressions of $S_N(t)$ for $\th=\pi/2$ and $Q=0$:
\begin{align}
&\left.  S_N(t)\right.= \frac{1}{N^N}   \sum_{\pp,\qq  }  \cos (E_{\pp,\qq} t) \, 
\det G(\pp) \det G(-\qq) \det F(\pp,\qq),
\label{eq:S(t)-2}\\
& 
F_{nm}(\pp,\qq)=  \frac {1}{\eE^{i(p_n-q_m)}-1}, \quad \quad n,m=1,2,\ldots N/2,    \nonumber \\
& G_{nm}(\pp)= 
  \eE^{2i n p_m} 
\left(1+ \eE^{-i p_m} \right),     \nonumber \\
&E_{\pp,\qq} =  4\sum_{j=1}^{N/2} (\cos p_j-\cos q_j). \nonumber
\end{align}
where $F,G$ are $\frac{N}{2} \times \frac{N}{2}$ matrices  and 
$\pp \equiv \{p_1,p_2, \ldots p_{N/2}\}$, $\qq \equiv \{p_1,p_2, \ldots p_{N/2}\}$, 
with $p_k$, $q_k$  all different and satisfying $e^{i p_k N}=1$,  $e^{i q_k N}=-1$,  
see \cite{2023ChiralBasis} for details.  
For $N=4,6$ (\ref{eq:S(t)-2}) gives
\begin{align}
&S_{4}(t) = \frac{1}{8} \left(2 \cos (4 t)+\left(3+2 \sqrt{2}\right) \cos \left(4 \left(\sqrt{2}-1\right)
   t\right)+\left(3-2 \sqrt{2}\right) \cos \left(4 \left(1+\sqrt{2}\right) t\right)\right) \nonumber \\
&S_{6}(t) =\frac{1}{96} \left(8 \cos (4 t)+2 \cos (8 t)+4 \cos \left(4 \sqrt{3} t\right)+\left(26+15
   \sqrt{3}\right) \cos \left(4 \left(\sqrt{3}-2\right) t\right)+\right.  \nonumber   \\
& +\left.  2 \left(7+4 \sqrt{3}\right) \cos
   \left(4 \left(\sqrt{3}-1\right) t\right)+2 \left(7-4 \sqrt{3}\right) \cos \left(4
   \left(1+\sqrt{3}\right) t\right)+\left(26-15 \sqrt{3}\right) \cos \left(4 \left(2+\sqrt{3}\right)
   t\right)+2\right).\nonumber
\end{align}

For sufficiently large $N$,  $S_N$ from Eq.(\ref{eq:S(t)-2}) gives an excellent approximation for  $S(t)$ up to times when $S(t)$ becomes vanishingly small,  see Fig.~\ref{Fig-S(t)}. The number of terms in $S_N$ (\ref{eq:S(t)-2}) grows exponentially with $N$.

\begin{figure}[tbp]
\centerline{
\includegraphics[width=0.5\textwidth]{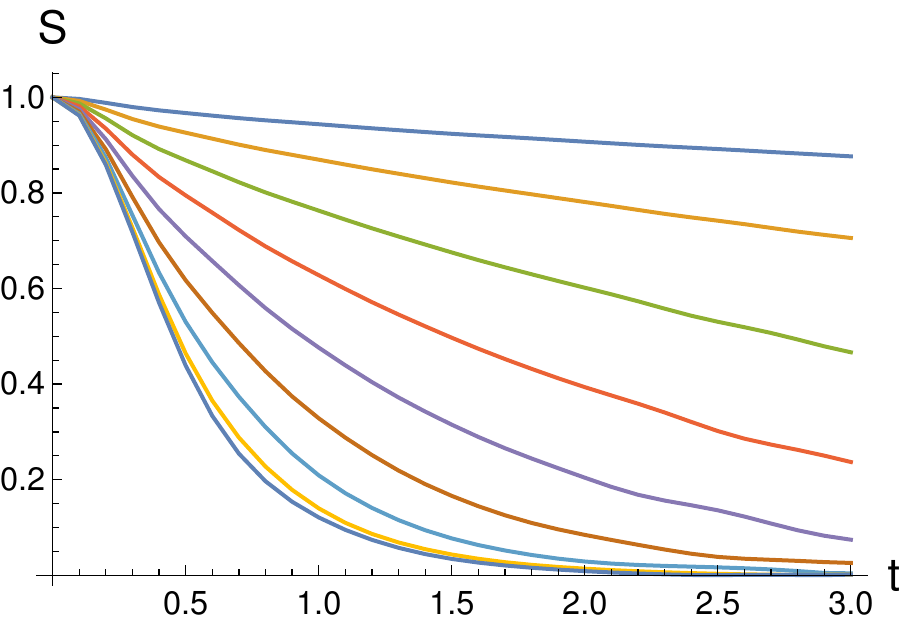}
}
\caption{
Decay in time of the rescaled amplitude $S(t)$ of homogeneous SHS $\ket{\Psi_0}$,  obtained from  the TEBD calculations for $N=64, \De=0$ and 
different $\th$.
Curves from top to bottom correspond to  increasing $\th$ values of $\th/\pi=0.1, 0.15,0.2, \ldots ,0.5 $.
}
\label{Fig-S(t)XX0}
\end{figure}

Eq.(\ref{eq:XX-Similarity}) allows to study just the homogeneous case $Q=0$ without losing generality. 
The curves $S(t)$ for $Q=0$ and different $\th$,  obtained via TEBD are given in Fig.~\ref{Fig-S(t)XX0}.  We see that 
the decay rate decreases with $\th$.  
 
For the two-point correlations $S^{\al,\be}_{n,m}(Q,t)= \langle\si_n^\al \si_m^\be(H_{XX},t)\rangle_Q$  we get for the 
simplest case $n=0$ using (\ref{eq:HxxTheorem}): 
\begin{align}
&S^{\al \pm}_{0,n}(Q,t) = \eE^{\pm in Q} \ S^{\al \pm}_{0,n}(0,  t \cos Q) \label{self-similarity2corr}\\
&S^{\al z}_{0,n}(Q,t) =  S^{\al z}_{0,n}(0,  t \cos Q). \label{self-similarity2corrZ}
\end{align}
For general $n,m$ the scaling relations can be obtained from  (\ref{self-similarity2corr}),(\ref{self-similarity2corrZ}) using (\ref{eq:shift}).
 Generalization  to the multipoint correlations is straightforward. 

\section{$XXZ$ case (arbitrary $\De$)}

For $\De \neq 0$ we find the early time evolution of observables using expansion 
\begin{align}
&e^X A e^{-X}  =A + [X,A] + \frac{1}{2!} [X,[X,A]] + \ldots = \sum_{n=0}^\infty \frac{1}{n!} ad^n_X(A) \label{eq:operatorExpansion}
\\
&ad_X(A)=[X,A], \quad ad^0_X(A) = A \nonumber
\end{align}
with $X$ substituted by $i H t$ and $A$ being the operator of a chosen observable.  For the expectation values,  (\ref{eq:operatorExpansion})
yields

\begin{align}
&\langle A(t) \rangle = \langle A(0) \rangle + \sum_{k>0} C_k \,t^k \label{eq:Taylor}
\end{align}
valid for the thermodynamic limit $N \rightarrow \infty$,  see Appendix~\ref{App:Taylor} for details.  
Generically,   a characteristic timescale  at which $\langle A(t) \rangle$ changes significantly can be estimated as

\begin{align}
&t_{char} = \left| \frac{C_{k_s}} {\langle A(0) \rangle}  \right|^{\frac{1}{k_s}}, \label{def:tchar}
\end{align}
where $C_{k_s}$ is the first nonvanishing coefficient in (\ref{eq:Taylor}).  For the rescaled SHS amplitude 
we obtain:
\begin{align}
&S(t) = 1 - 4 t^2 (\De - \cos Q)^2 \sin^2 \th + O(t^4) = 1-\left(  \frac{t}{t_{char}} \right)^2 + O(t^4)
\label{eq:TaylorSHS}\\
&t_{char} = | 2 (\De - \cos Q) \sin \th|^{-1}. \label{eq:tchar}
\end{align}
%$t_{char}$ roughly estimates the time needed for SHS amplitude to decay to zero,  restoring $U(1)-$ symmetry of the $XXZ$ Hamiltonian
%for one-site reduced density matrices
%as discussed in Eq. (\ref{eq:commU(1)-operators}).

So,  $S(t)$ initially decreases with time,  thus driving the $1$-site density matrix towards $U(1)$ -symmetric point  $S=0$.
The states with larger initial  amplitude
(larger $\sin \th$) decay faster.  The amplitude expansion (\ref{eq:TaylorSHS}) for $\De=0$ up to the order
$t^{12}$ is given in  the Appendix~\ref{App:Taylor}.

\begin{figure}[tbp]
\centerline{
\includegraphics[width=0.5\textwidth]{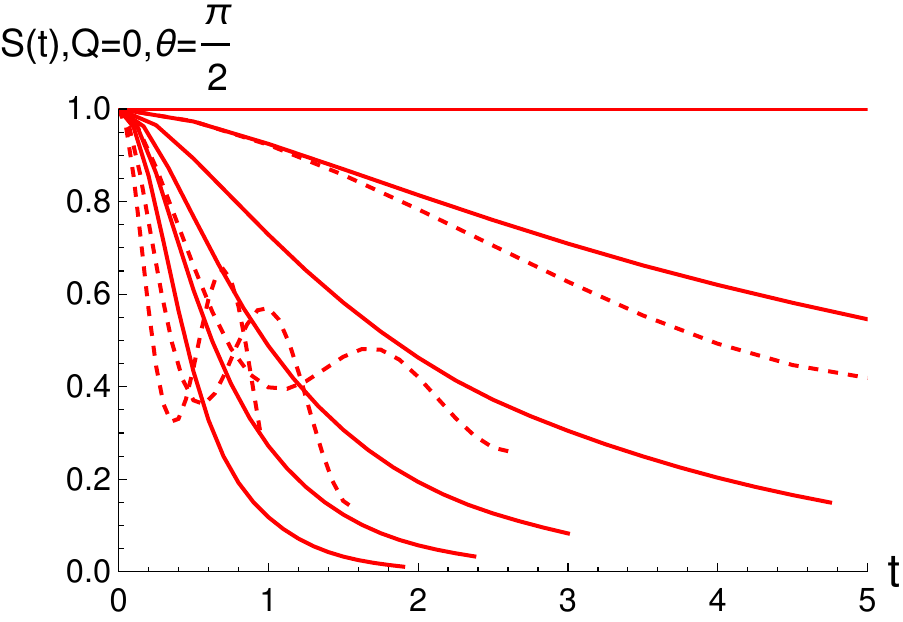}
\includegraphics[width=0.5\textwidth]{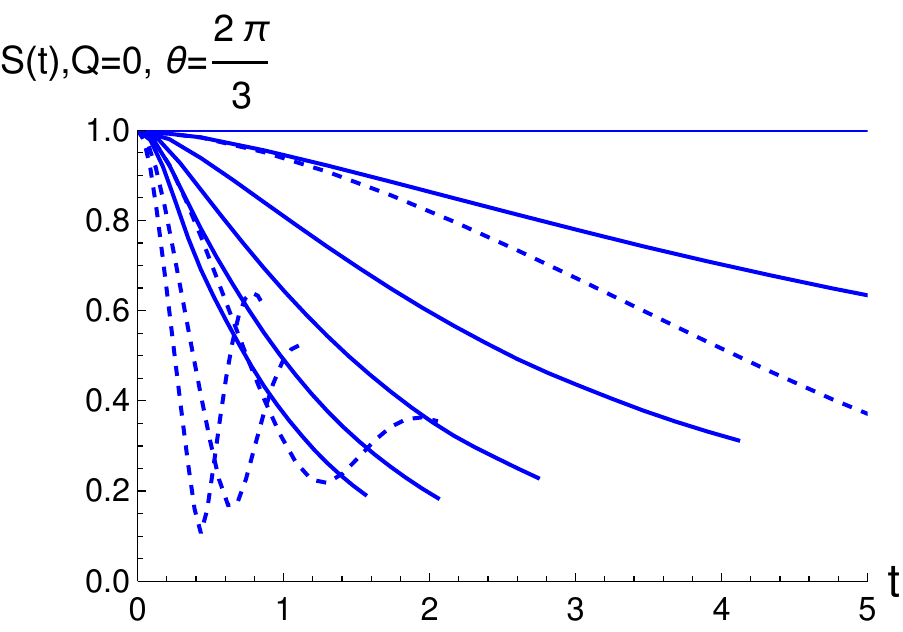}
}
\caption{
Rescaled SHS  amplitude $S(t)$ for $Q=0$ for different $\De$,  for $\th=\pi/2$ (Left Panel) and 
 $\th=2\pi/3$ (Right Panel), from TEBD for open $XXZ$ spin $1/2$ chain with $64$ sites. 
Full curves correspond to $|\De|\leq 1$, namely $\De = 1,0.8, 0.6,0.4,0.2, 0$ (from top to bottom).  
Dashed curves correspond to $\De>1$,  namely
$\De=1.2,1.8,2.4,3$,  the shorter 
curves (i.e.  those  with smaller maximal value of $t$ shown) correspond to larger $\De$. 
The largest reported time for each dashed curve is given by $3 \, t_{char}$ from (\ref{eq:tchar}).
%Right Panel: dashed line denotes the threshold value $\cos^2 \th$ for $S(t)$ drop in the $\De \rightarrow \infty$ limit. 
}
\label{Fig-S(t)-De}
\end{figure}

Another important feature of  (\ref{eq:tchar}) is the divergence of $t_{char}$ for $\De \rightarrow \cos Q$.  Indeed for  $\cos Q=\De$
the SHS $\ket{\Psi_{Q,\th,\vfi}}$ becomes an $XXZ$ eigenstate \cite{SHS-Phantom},    leading 
to the time-independent  SHS amplitude,  and the $t_{char}\rightarrow \infty$.
To obtain $S(t)$ for intermediate times $t=O(1)$ and arbirary $\De$ we use TEBD. 
 Fig.~\ref{Fig-S(t)-De} shows the rescaled amplitude $S(t)$ for $Q=0$ and different values of $\De$ and $\th$.

\begin{figure}[tbp]
\centerline{
\includegraphics[width=0.5\textwidth]{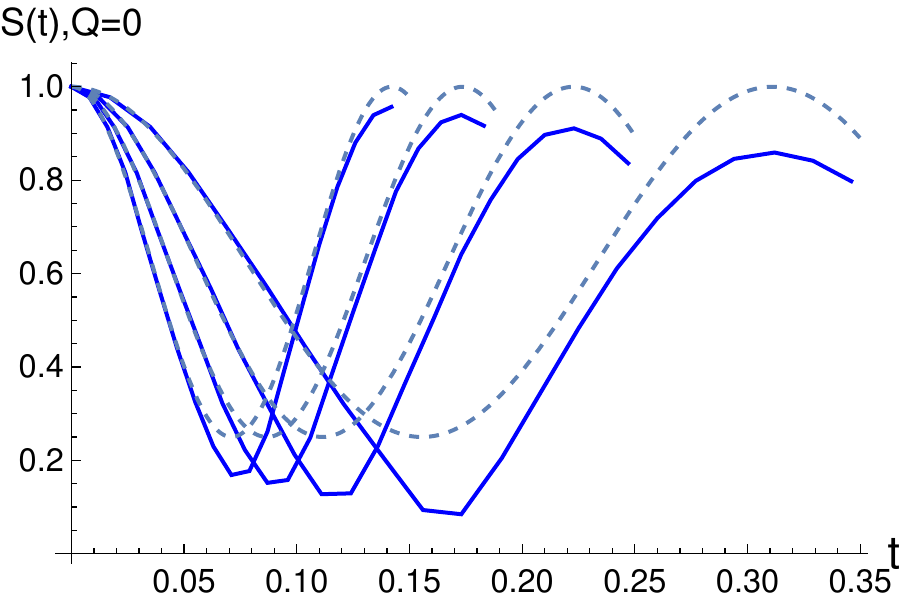}
}
\caption{
Rescaled SHS  amplitude $S(t)$ for $Q=0$ for large $\De$,  for $\th=2\pi/3$ from TEBD.
Full curves connecting datapoints correspond to $\De=6,8,10,12$, while 
 dashed curves show theoretical $\De\gg 1$ expressions (\ref{S(t)DeLarge})
 with fitted non-universal shift $a=0.9$.
Shorter curves correspond to larger $\De$. 
}
\label{Fig-DeLarge}
\end{figure}

We observe the onset of oscillations of $S(t)$ for $\De > 1$ (dashed curves in  Fig. ~\ref{Fig-S(t)-De} )
with slower overall decay of the oscillation envelope. 
The emergence of the oscillations can be understood by considering $\De\gg 1$ limit,
which yields an oscillatory solution for $S(t)$ (see Appendix \ref{App:Ising}):
\begin{align}
&\left. S(t) \right|_{\De \gg 1} = \frac{1+\cos^2 \th} {2} + \frac12 \sin^2 \th \ \cos (4 (\De-a) t), \label{S(t)DeLarge}
%& \min_t \left. S(t) \right|_{\De \gg 1} = \cos^2 \th \label{eq:S(t)min}
\end{align}
where non-universal shift $a=O(1)$ results from $1/\De$ corrections.  In addition,  the presence of hopping terms (neglected in the derivation of 
(\ref{S(t)DeLarge})) leads to overall decay of the oscillation envelope with time, well visible in the TEBD data for $\De \gg 1$, see Fig.~\ref{Fig-DeLarge}.

%\min_t \left. S(t) \right|_{\De \gg 1} = \cos^2 \th \label{eq:S(t)min}

Following \cite{SHS-Ketterle},
we shall characterize the SHS decay  $S(t)$ by a parameter $\ga$,  the inverse time of the amplitude decay by one-half,
\begin{align}
&S (1/\ga) = \frac{S(0)}{2} = \frac12.\label{def:ga}
\end{align}
A rough estimate  for $\ga$ is given by  the quantity $1/t_{char}$ in (\ref{eq:tchar}).  
For monotonic $S(t)$ dependence as in Fig.~\ref{Fig-S(t)XX0},  $\ga$ has the meaning of a decay rate of the amplitude,  while for 
$\De \gg 1$ when (\ref{S(t)DeLarge}) is approximately valid,  $\ga$ gives a measure for the rate of the amplitude change.    

Note that the asymptotic $\De \rightarrow \infty$ values of $S(t)$ (\ref{S(t)DeLarge}) obey $ S(t) \geq \cos^2 \th$.  Our definition of $\ga$ in (\ref{def:ga}) is therefore valid for  $\cos^2 \th <1/2$, i.e. when  $\pi/4 < \th < 3 \pi/4$.
 From (\ref{S(t)DeLarge}) we obtain:
\begin{align}
&\left. \ga \right|_{\De \gg 1} = \frac{8 (\De-a)}{\pi + 2 \arcsin(\cot^2 \th )}, \nonumber
\end{align}
predicting the asymptotic slope 
\begin{align}
&\left. \frac{\partial \ga}{\partial \De} \right|_{\De \gg 1} = \frac{8 }{\pi + 2 \arcsin(\cot^2 \th )} \label{eq:gaDeLargeSlope}
\end{align}

\begin{figure}[tbp]
\centerline{
\includegraphics[width=0.5\textwidth]{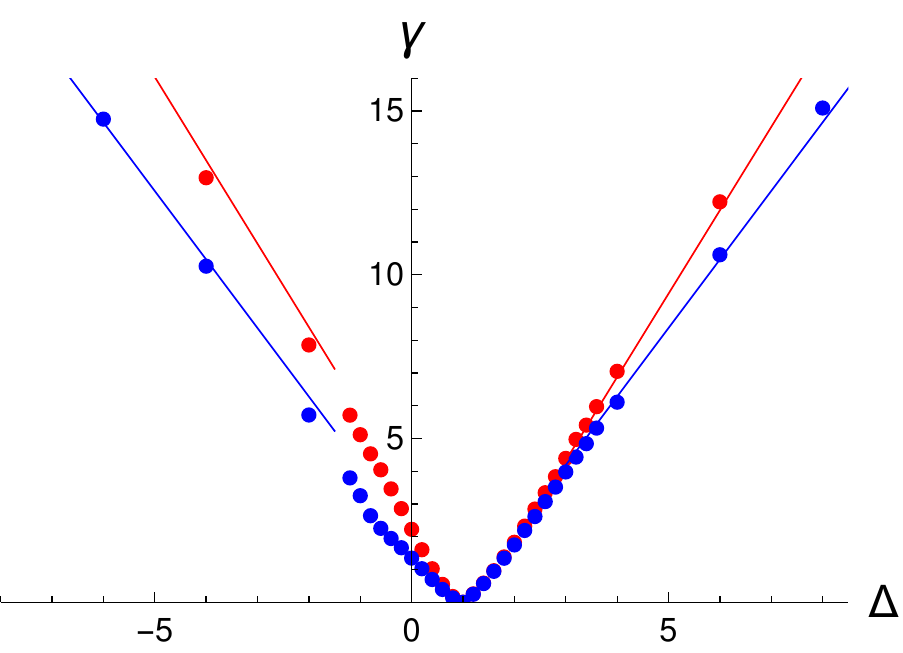}
\includegraphics[width=0.5\textwidth]{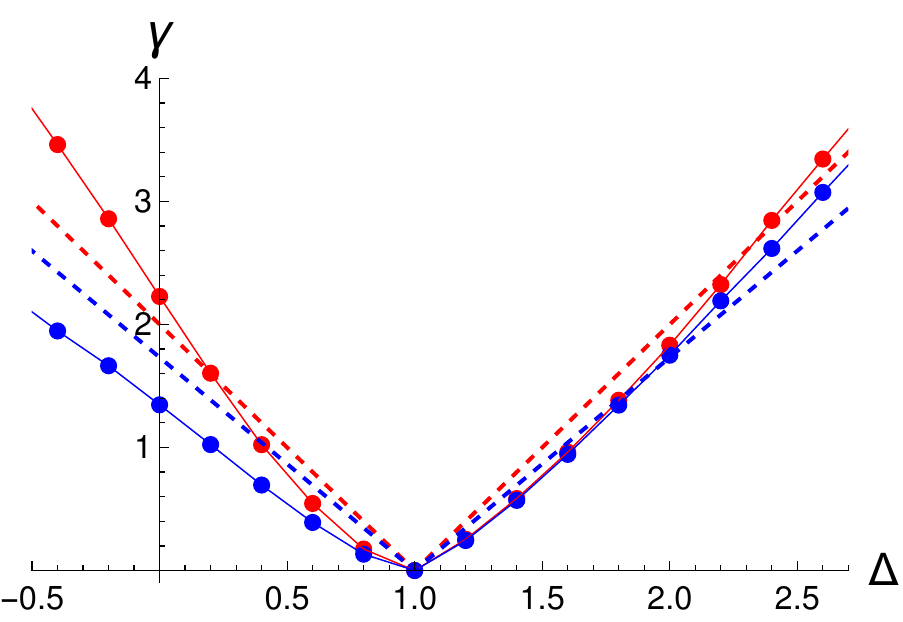}
}
\caption{
Half-amplitude decay rate $\ga$ versus $\De$ for $Q=0$ and $\th=\pi/2$ (red dataset),
$\th=2\pi/3$ (lower blue dataset ), calculated from the data similar to the ones shown Fig.~\ref{Fig-S(t)-De}, respecting the colour code. Right Panel is a closeup of the Left Panel.
\textit{Left Panel}: Full straight lines have slopes, predicted from (\ref{eq:gaDeLargeSlope}) for the $\De \rightarrow \infty$ limit. 
\textit{Right Panel}: Dashed straight lines show $t_{char}^{-1}=2| (\De - 1) \sin \th|$ for  $\th=\pi/2$ (red),
$\th=2\pi/3$ (blue). 
}
\label{Fig-gamma}
\end{figure}

The decay rates $\ga$ versus $\De$ for $Q=0$ and fixed $\th$,  shown  in Fig.~\ref{Fig-gamma},
play a crucial role. 
In what follows we demonstrate that  $\ga$ for arbitrary  $Q$ is obtainable from 
$\ga$ for $Q=0$ via an approximate scaling.

\section{Approximate scaling for the SHS amplitude decay rate }

It is tempting to generalize convenient  scaling property (\ref{eq:HxxTheorem}) relating expectations taken with respect to
homogeneous ($Q=0$)  and nonhomogeneous  ($Q\neq 0$)  initial state  for $\De \neq 0$ regime. 
 Simply ``ignoring" the $J$-containing Dzyaloshinskii-Moriya (DM) term in (\ref{eq:H'}) is not possible: even if  in the first order in time,  the DM term gives no contribution 
\begin{align*}
& e^{-i H't}\ket {\Psi_{0}}\approx (I-i t\, H'  + O(t^2))\ket {\Psi_{0}}\\
&=\left(I -it  \left.  H'\right|_{J\rightarrow 0}  + O(t^2)\right) \ket {\Psi_{0}},
\end{align*}
already for $t^2$ order the DM term cannot be neglected,  since $J$ and $H$ do not commute.  

\begin{figure}[tbp]
\centerline{
\includegraphics[width=0.5\textwidth]{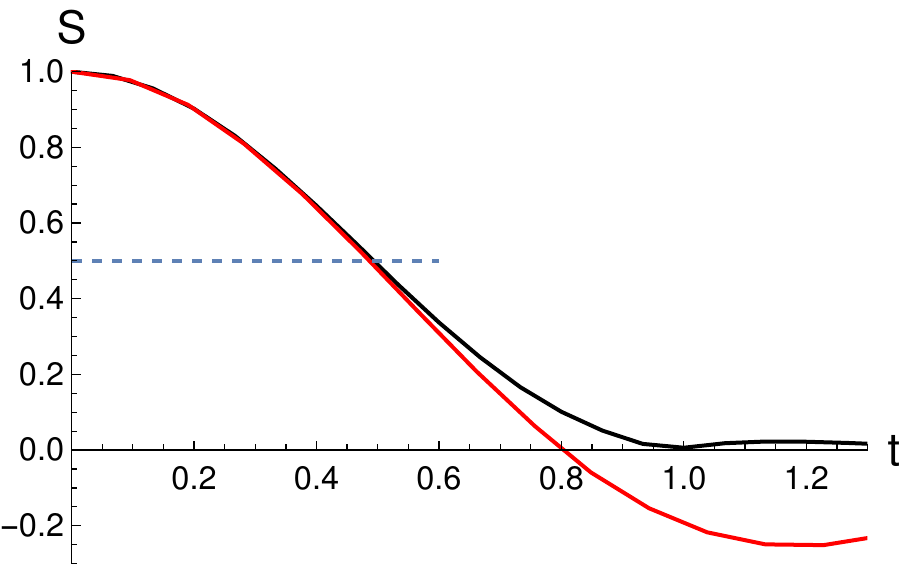}
}
\caption{
Illustrating early time equivalence of $S(Q,\De,t)$(black curve)  and $S(0,\De/\cos Q, t \cos Q)$ (red curve)
 for $\De=0.5, Q= 1.87 $.  Despite being qualitatively different at later times,  the two curves 
overlap at   early times,  which
allows an accurate estimate of $\ga$ from  (\ref{eq:gaDeScaling}) at the threshold value given by crossing of $S(t)$
with the dashed line,  explaining the data collapse in  Fig.~\ref{Fig-DataCollapse}.
}
\label{Fig-S(t)Comparison}
\end{figure}

\begin{figure}[tbp]
\centerline{
\includegraphics[width=0.5\textwidth]{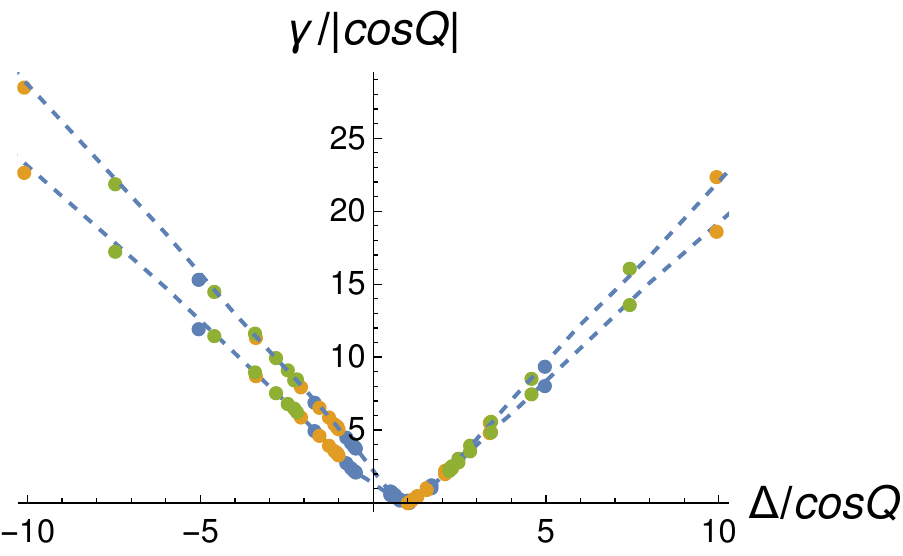}
}
\caption{
Data collapse.  Rescaled  $\ga$ are plotted versus rescaled $\De$ for set of chosen parameters: $\De=0.5,1,2.2$(blue,yellow, green points respectively) and equidistant $Q=0.07, 0.27,  \ldots 3.07$, 
for $\th= \pi/2$ (upper dataset),  $\th= 2\pi/3$ (lower  dataset).  Dashed curves show $\ga(0,\De)$ for $Q=0$ obtained 
from the points  in Fig.~\ref{Fig-gamma} by linear interpolation.  
}
\label{Fig-DataCollapse}
\end{figure}

However in the expansion (\ref{eq:Taylor})
some coefficients $C_k$,  $k\geq 2$ can become $J$- independent,  depending on the observable.   In particular,
one-point correlations $\langle \si_0^\al \rangle$ turn out to be   $J$- independent up to the order $t^5$! 
E.g.  for $\th = \pi/2$  we obtain
\begin{align}
&\langle\si_0^x(H,t) \rangle_Q = 1- \ka(Q) t^2 + \frac{4 t^4}{3} \ka(Q)\, (2 \cos^2 Q+\De^2 - \De \cos Q ) + O(t^6) 
\label{eq:SigmaxTaylor}\\
&\ka(Q) = 4(\De-\cos Q)^2 \nonumber
\end{align}
and the same Taylor expansion is valid for $\langle \si_0^x(H|_{\De \rightarrow \De/\cos Q},\,t\cos Q)\rangle_0$,  
 as can easily checked !
Equivalence of  $S(Q,\De,t) -  S(0,\frac{\De}{\cos Q}, t \cos Q) = O(t^6)$ means
approximate overlap of the two functions at early times,    
and in most cases the overlapping region extends up to the time of one-half amplitude decay,  as exemplified in
Fig.~\ref{Fig-S(t)Comparison}.  
This allows to  relate the half-amplitude decay rates $\ga$ of homogeneous and non-homogeneous SHS in an approximate way via
\begin{align}
&\ga(Q,\De) \approx \cos Q \times \ga(0, \frac{\De}{\cos Q}), \label{eq:gaDeScaling}
\end{align}
generalizing exact (\ref{eq:XX-Similarity}) for $\De \neq 0$.
Technically,  (\ref{eq:gaDeScaling}) predicts the data 
collapse of $\ga$ points with rescaled coordinates 
$\{\De/\cos Q,  \ga(Q,\De)/|\cos Q|  \}$ on a single curve $\{\De,   \ga(0,\De)  \}$,  i.e.  the on curves  Fig.~\ref{Fig-gamma}.  
Such a data collapse indeed exists,  see Fig.~\ref{Fig-DataCollapse}, and it  confirms the validity of (\ref{eq:gaDeScaling}).
In addition,  the approximate relation (\ref{eq:gaDeScaling}) becomes exact in various  limits: $\De=0$,  $\De \rightarrow \cos Q$,  $Q \rightarrow 0,\pi$,  and also
for $\De \gg 1$,  up to $\De^{-1}$ corrections.  Note that $\ga(Q,\De)$ for some polar angle $\th$ is related by (\ref{eq:gaDeScaling}) to universal 
curve $\ga(0,\De)$ with the same value of $\th$.   The  curve   $\ga(0,\De)$ at large $\De$ becomes a straight line with slope (\ref{eq:gaDeLargeSlope}).

\begin{figure}[tbp]
\centerline{
\includegraphics[width=0.55\textwidth]{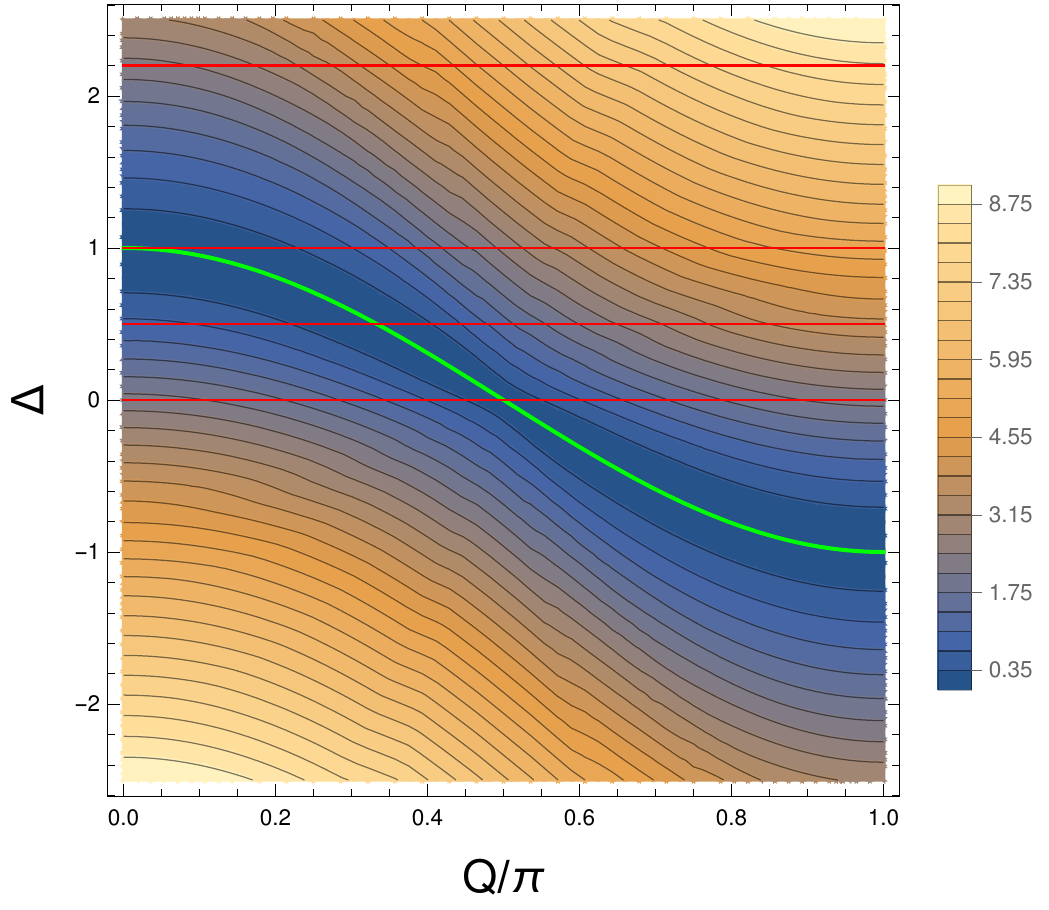}
}
\caption{
Contour plot of decay rate $\ga$ versus $Q/\pi$ and $\De$,  for $\th=\pi/2$,  obtained by using scaling relation (\ref{eq:gaDeScaling}). 
The green curve  $\De=\cos Q$, where the SHS eigenvalue condition is met,  gives the location of points with no decay 
$\ga=0$.  Cuts 
of the surface $\ga(Q,\De)$  alond the red straight lines are shown in red in Fig. ~\ref{Fig-gamma(Q)}
}
\label{FigContourGamma}
\end{figure}

We conclude that the half-amplitude decay rate  $\ga$ for nonzero $Q$ can be found with good accuracy by measuring the decay rate of a homogeneous initial state with the same value of $\th$   and a rescaled anisotropy $\De \rightarrow \De/\cos Q$,  via  (\ref{eq:gaDeScaling}). 

\begin{figure}[tbp]
\centerline{
\includegraphics[width=0.45\textwidth]{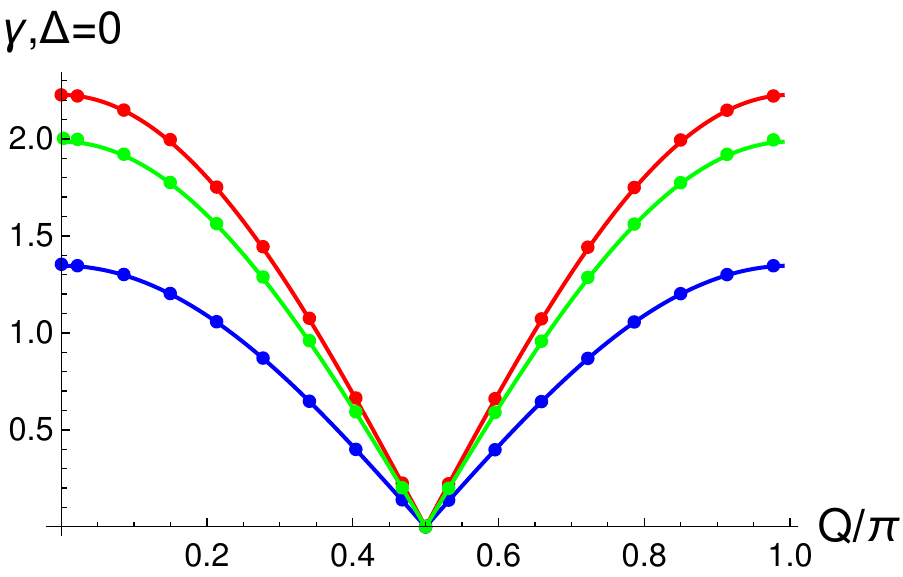}
\includegraphics[width=0.45\textwidth]{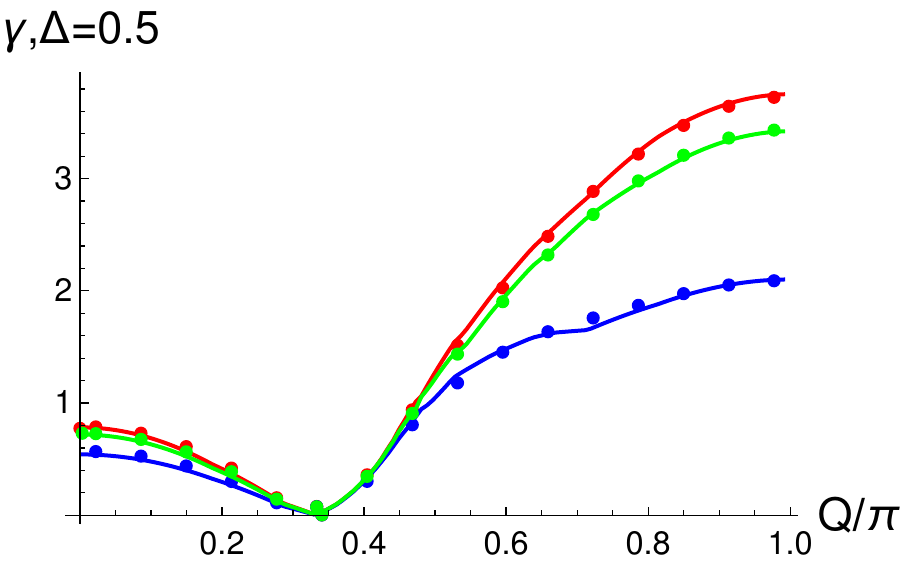}
}
\centerline{
\includegraphics[width=0.45\textwidth]{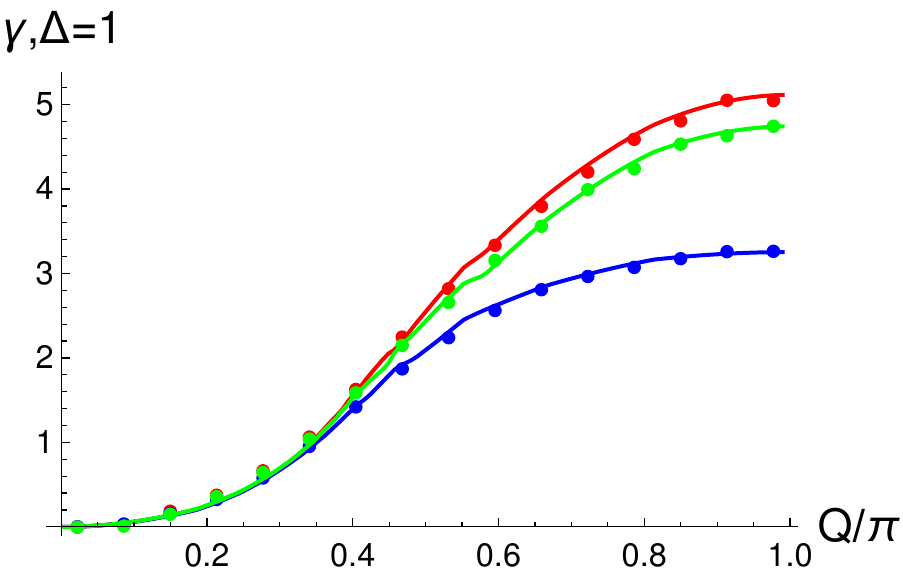}
\includegraphics[width=0.45\textwidth]{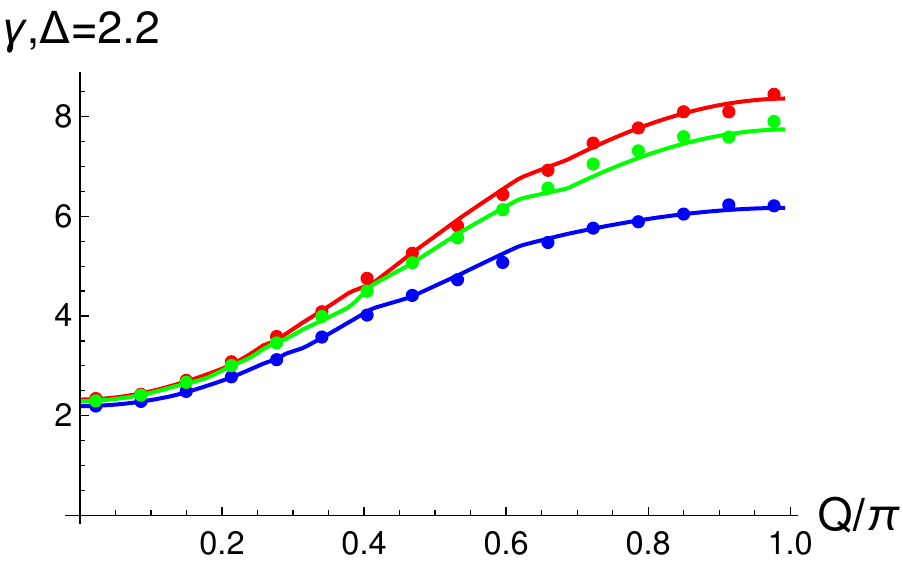}
}
\caption{ 
Decay rate $\ga$ versus $Q/\pi$ for fixed $\De$,  reported on $y$- axis label,  and selected $\th = \frac{\pi}{2},  \frac{5 \pi}{12},
\frac{2 \pi}{3} $ (red, green, blue  datapoints respectively).
 Curves with the same colour code show $\ga$ calculated using the scaling relation (\ref{eq:gaDeScaling})
and numerical data for $Q=0$ ( like that in  Fig.\ref{Fig-gamma} for $\th = \pi/2$ and $\th=2 \pi/3$). 
}
\label{Fig-gamma(Q)}
\end{figure}

Now we can generate the decay curves and make comparison to experimental data. 
By linear extrapolation of TEBD data from  Fig.~\ref{Fig-gamma} on the full $\De$ axis, we obtain 
a set of curves $\ga(0,\De)$ for different $\th$.
From the curves $\ga(0,\De)$ we generate surface $\ga(Q,\De)$ (see Fig.~\ref{FigContourGamma})  using  (\ref{eq:gaDeScaling}),  and compare 
few cuts of this surface with the direct TEBD 
data (Fig.~\ref{Fig-gamma(Q)} ) and with the  experimental data for the same quantity, called decay of contrast, 
 in \cite{SHS-Ketterle}.  
The data Fig.~\ref{Fig-gamma(Q)}  confirm the validity of (\ref{eq:gaDeScaling}),   and
 are in qualitative accordance with the experimental data ( Figs.3a-3c  in \cite{SHS-Ketterle}) 
as well but there are also  discrepancies  worth discussing. 
The fit-function $\ga(Q) = \ga_0 + \ga_1 |\De - \cos Q|$, used in \cite{SHS-Ketterle}
to  find the anisotropy $\De$, is equivalent to an approximation $\ga = const *(1/t_{char})$  where $t_{char}$ is given in (\ref{eq:tchar}),
if the offset $\ga_0$ ( due to noise) is neglected,  $\ga_0 \rightarrow 0$ (another fit function used in \cite{SHS-Ketterle},  $S(t)=1-\ga t$, violates $S(t)=S(-t)$ symmetry).  While for $\De=0$, the approximation 
$\ga \sim t_{char}^{-1}$  is exact due to 
 (\ref{eq:XX-Similarity}), and is reasonable for large $\De$ (where $\ga$ is also large and 
 $Q$-dependence can be neglected, see  (\ref{S(t)DeLarge})), 
for  $\De \neq 0$ and  intermediate $\ga$ values it appears rather poor,  see discrepancy between dashed and full lines 
in the Right Panel of Fig.~\ref{Fig-gamma}.
A usage of scaling relation (\ref{eq:gaDeScaling}) with tabulated $\ga(0,\De)$ appear  more promising  option for 
the calibration of the anisotropy.

\section{SHS: phase velocity of an asymptotic travelling wave}
\label{sec:PhaseVelocity}

\begin{figure}[tbp]
\centerline{
\includegraphics[width=0.5\textwidth]{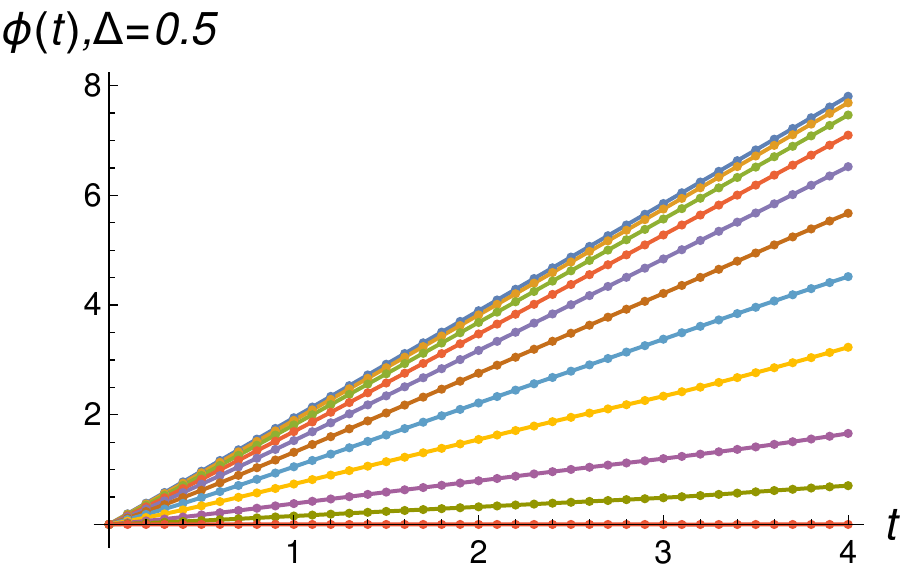}
\includegraphics[width=0.5\textwidth]{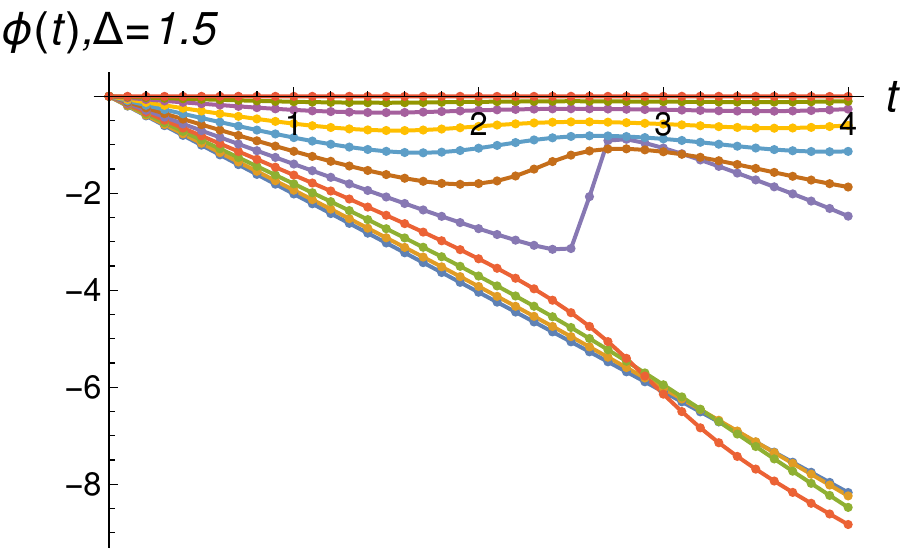}
}
\caption{
Datapoints   show phase $\phi$   versus  time, obtained by tracking position of constant 
phase in the decaying SHS (see green line in Fig. ~\ref{Fig-TravellingWave}),
from TEBD calculations, for $\De=0.5$ (left Panel)
 and   $\De=1.5$ (right Panel), for 
 $Q=0.14$ and different polar angles $\th$. Curves from top to bottom in the left Panel 
(bottom to top in the right Panel)
correspond to  increasing $\th$ values  $\th/\pi=0.05,0.1, 0.15,0.2, \ldots ,0.5 $. 
}
\label{Fig-Phase}
\end{figure}

\begin{figure}[tbp]
\centerline{
\includegraphics[width=0.5\textwidth]{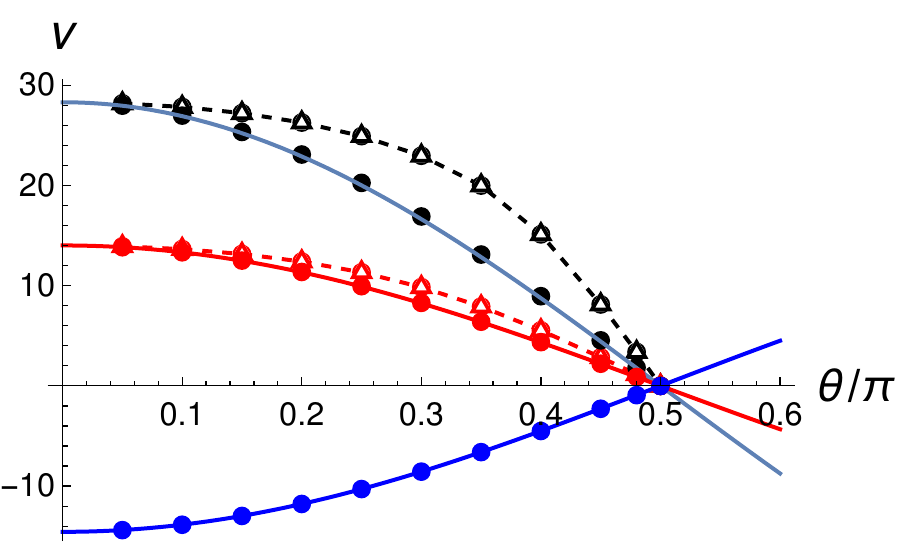}
}
\caption{
Phase  velocity of decaying SHS $v=\phi'(t)/Q$, for  $t=0$ (filled symbols) and $t=t_{max}$ (open symbols) ,  versus $\th/\pi$,  obtained from  the TEBD calculations,  for $Q=0.14$,
 $\th = \frac{\pi}{2}$ and different anisotropies
$\De =0,0.5, 1.5$ (black, red and blue points, respectively).  Full lines show the prediction  (\ref{eq:PhaseVelocity}).
Open symbols show the asymptotic phase velocity, determined from TEBD,
the dashed lines are guides for an eye.  For $\De=1.5$ (the blue data) the asymptotic phase velocity cannot be 
determined  because of oscillatory behaviour,  see text.  }
\label{Fig-PhaseVelocity}
\end{figure}

Finally,  we discuss the behaviour of the SHS phase $\phi(t)$ in (\ref{eq:SigmaX}),(\ref{eq:SigmaY}).  From (\ref{eq:parityS}), (\ref{eq:vThetaDep}), 
$\phi(t)$ is an odd function of $t$ and is nonzero only for 
$\th\neq \pi/2$,  and for $\De \neq \cos Q$.   For the easy plane regime $|\De|  \leq  1$,  $\phi(t)$  is an approximately linear function of $t$ 
(see data in Fig.\ref{Fig-Phase} , left Panel),
thus allowing to view the SHS profile  evolution as a  decaying travelling wave with time-dependent phase velocity
as in Fig.~\ref{Fig-TravellingWave}.  Indeed,  for  linear $\phi(t)$  the  (\ref{eq:SigmaX}) becomes
\begin{align}
&\frac{\langle \si_n^x(t) \rangle}{\sin \th}  =  S(t)  \cos(Q (n -v t) +\vfi ), \label{eq:SigmaX-asymptotic}\\
& v= \frac{1}{Q} \frac{\partial \phi }{\partial t}. \nonumber
\end{align}
The  initial phase velocity $v_0$ at time $t=0$ can be 
estimated by a perturbative  analysis (treating transversal SHS components as a perturbation,  see  Appendix~\ref{App:v(0)}), 
yielding 
 \begin{align}
&\left.  
\frac{1}{Q}\frac{\partial \phi }{\partial t}  \right|_{t=0}  =  \,v_0     =   \frac{4(\cos Q - \Delta) \cos\theta}{Q},\label{eq:PhaseVelocity}
\end{align} 
which we find in qualitative accordance with numerics, see full lines in Fig.~\ref{Fig-PhaseVelocity}.
For small $\th $ (or $\pi-\th$) values,  the phase velocity (\ref{eq:PhaseVelocity})
stays constant in time,  see Appendix \ref{Theta0}.  
Generically, for  $|\De|  \leq  1$,  the phase velocity $v=\phi'(t)/Q$ changes monotonically in time and quickly reaches an asymptotic value,  which can be determined numerically, see
open symbols in Fig.~\ref{Fig-PhaseVelocity}.
On the other hand,   in the easy axis regime ($|\De|>1$)   
$\phi'(t)$ shows decaying oscillatory behaviour,
see right Panel of Fig.~\ref{Fig-Phase}, related to the  oscillatory behaviour of $S(t)$ itself,
see dashed curves in  Fig.~\ref{Fig-S(t)-De}.  We expect that an asymptotic phase velocity also exists in this case but
it  cannot be determined from TEBD  because of slow convergence.  

\section{Time dependence of magnetization current under  the $XXZ$ dynamics }
At the end,  we discuss the time evolution of the SHS magnetization current  $j(t) \equiv \langle j^z (t)\rangle$.  
Operator $j^z$ is a $U(1)$-invariant quantity,  and therefore $j(t) $ is not expected to decay  with time. 
From  (\ref{eq:operatorExpansion}) we find 
the early time behavior 
\begin{align}
& j(t)=j(0) -  8 t^2 \De \,\sin Q  (\De-\cos Q) \, \sin^4 \th + O(t^4),\label{eq:jzExpansion}
\end{align}
where $j(0)= 2 \sin Q \sin^2 \th$ is the initial SHS magnetization current.  From  (\ref{def:tchar}) the  characteristic timescale for the current  to change is given by 
\begin{align}
&t_{char}(J)  =\left| {2 \De (\De - \cos Q) \sin^2 \th} \right|^{-\frac12}. 
\label{eq:tcharCurr}
\end{align}
For $\De=0$,  the magnetization current is a conserved quantity,  so it is time-independent which is reflected in the divergence of $t_{char}(J) $ at $\De=0$.

\begin{figure}[tbp]
\centerline{
\includegraphics[width=0.5\textwidth]{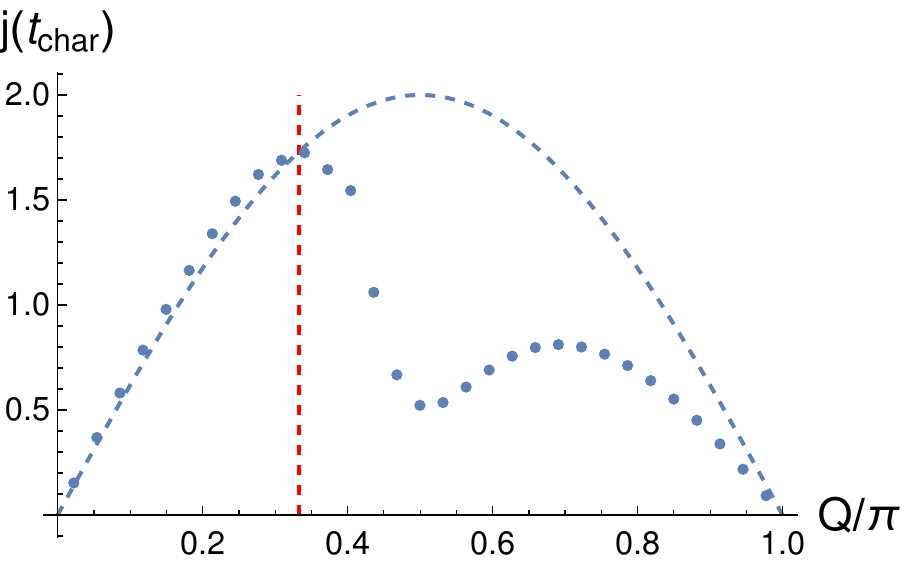}
}
\caption{
Magnetization current $j$ measured at time $t_{char}$  (\ref{eq:tcharCurr}),  from TEBD (points), versus 
$Q/\pi$.
Parameters: $\De=0.5, \th = \pi/2$. Dashed curve shows the initial SHS current $ j(0)  = 2 \sin Q$.
Dashed vertical line at $Q=\pi/3$ corresponds to special point $\cos Q=\De$ for which $\ket{\Psi_Q}$ is an $H$ eigenstate and the current stays constant in time. 
}
\label{Fig-J}
\end{figure}

\begin{figure}[tbp]
\centerline{
\includegraphics[width=0.5\textwidth]{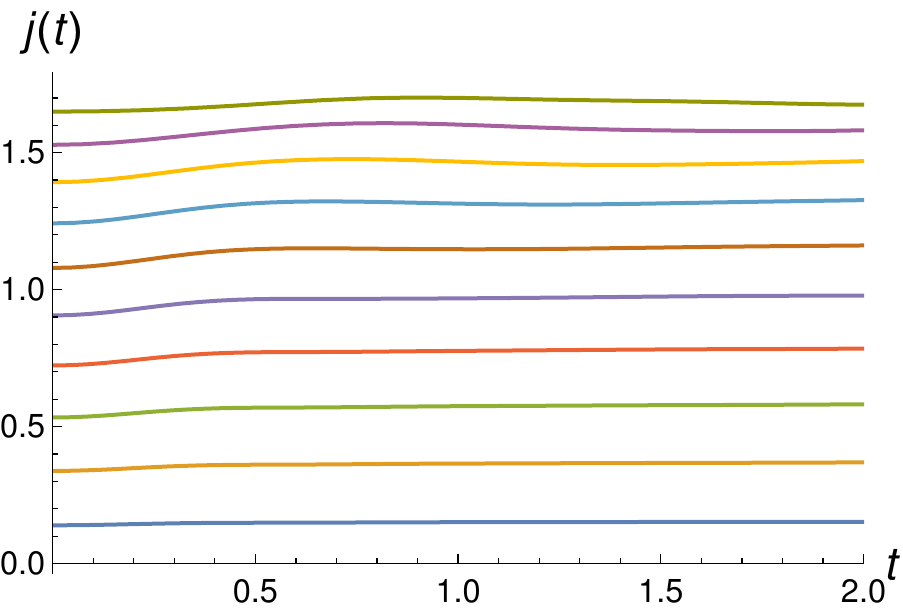}
\includegraphics[width=0.5\textwidth]{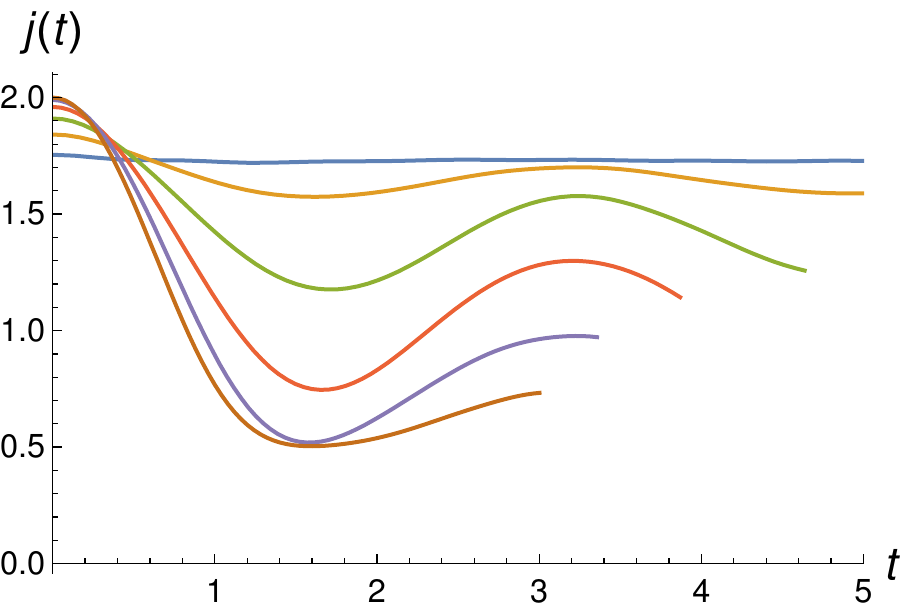}
}
\caption{
Magnetization current $ j(t) $,  for different $Q$,  for $\cos Q >\De=0.5$ (Left Panel) and 
 $0 \leq \cos Q \leq \De=0.5$ (Right Panel), from TEBD. 
Parameters Left Panel: $\th=\pi/2$, $Q=0.07, 0.17, \ldots 0.97$ (curves from bottom to top). 
Parameters Right Panel: $\th=\pi/2$, $Q=1.07, 1.17, \ldots ,1.57\approx \pi/2$ (curves from  top to bottom at the right corner). The largest reported time for curves at the Right Panel  is given by $3 \, t_{char}(J)$ (\ref{eq:tcharCurr}).
}
\label{Fig-J(t)-Q}
\end{figure}

From the TEBD, we observe that 
that the early time behaviour  sets the gradient current at later times,  namely,  
\begin{align}
& j^z(t>0) -  j^z(0) > 0  , \quad \mbox {if $\De (\cos Q-\De) \sin Q >0 $},   \label{eq:signAsymCurrent}\\
& j^z(t>0) -  j^z(0) < 0  ,\quad  \mbox {if $\De (\cos Q-\De) \sin Q <0 $},  \nonumber
\end{align}
see Fig.~\ref{Fig-J(t)-Q} for an illustration. 
The TEBD data for $\De=0.5$  (Fig.~\ref{Fig-J}) and for other $\De$ values (data not shown)   suggest  (a) validity of (\ref{eq:signAsymCurrent}) (b) qualitative change of behaviour 
$ j(t)$ across the eigenfunction SHS point $\cos Q = \De$, exemplified by the difference between the curves on the Left and on the Right Panel in 
Fig.~\ref{Fig-J(t)-Q}. (c) the wavevector $Q=\pi/2$ (where the $t=0$ SHS current is maximal) becomes a local minimum 
for the  current value at late times.  

The observations (a) - (c) deserve further study.  Especially it would be interesting to 
see if (a)-(c) are valid also for the asymptotic value of the current $ j(\infty)$.
Unfortunately,  we cannot determine $ j(\infty)$
precisely enough because of the fast grows of the entanglement with time,  and slow convergence.
For selected parameters we tried to push the TEBD calculations as far as possible, however the ``entanglement'' barrier 
  prevents precise measurements for $t>4$, which is not enough to see the convergence, see Fig.~\ref{Fig-Marko}, and 
 technical details in Appendix \ref{App:DMRG}. 

%see Fig.~\ref{Fig-jz}

\begin{figure}[tbp]
\centerline{
\includegraphics[width=0.5\textwidth]{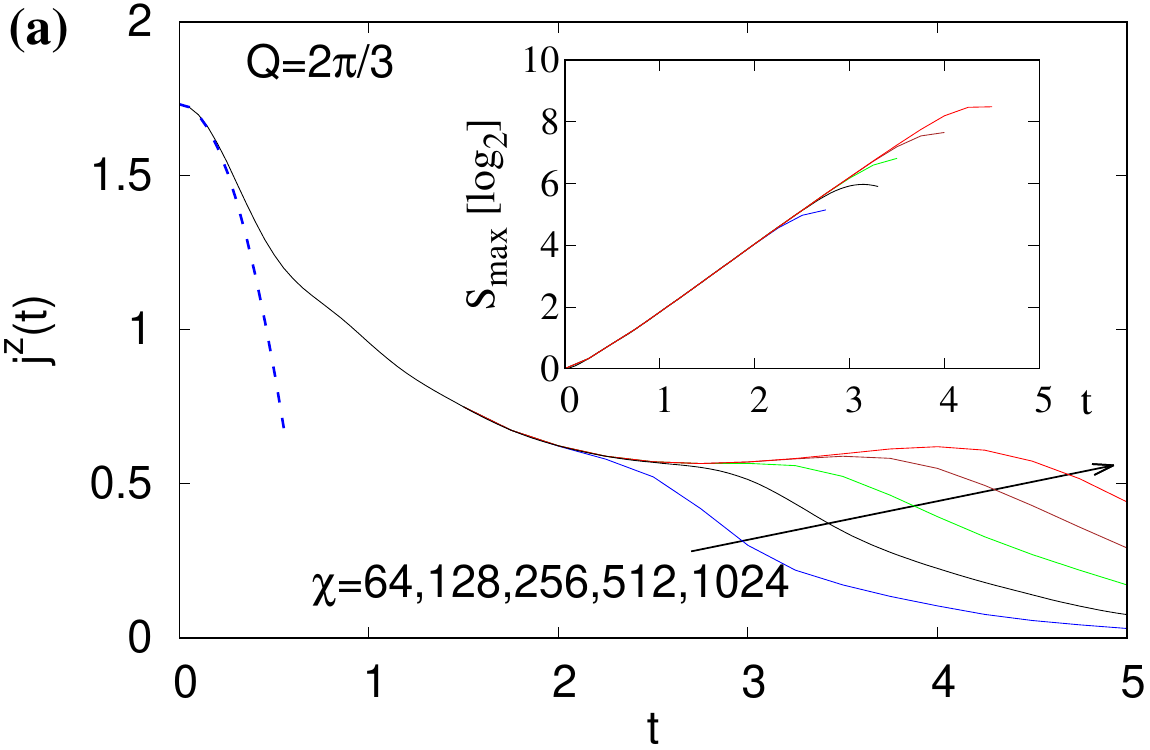}
\includegraphics[width=0.5\textwidth]{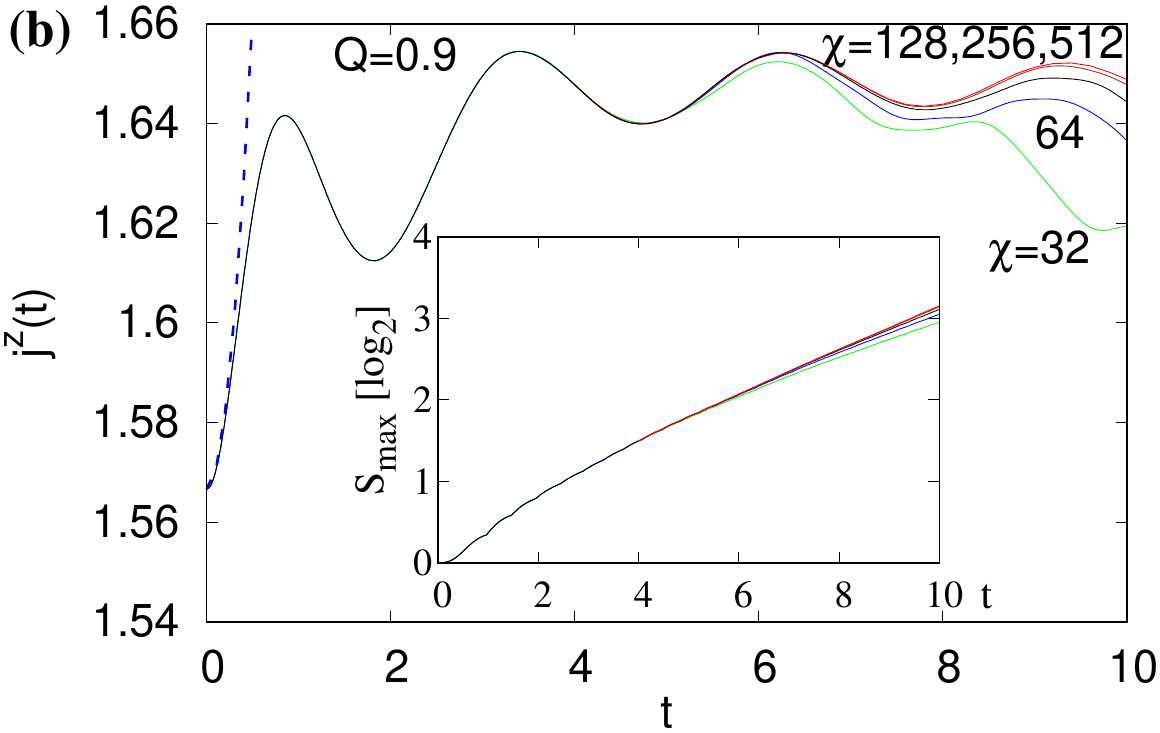}
}
\caption{
Magnetization current $\langle j^z \rangle $   versus  time , 
from TEBD calculations with varied size $\chi$ of MPS matrices in the TEBD method, see Appendix~\ref{App:DMRG} for 
details.  
Parameters Left Panel: $\De=0.5, Q=2\pi/3, \th=\pi/2$. 
Parameters Right Panel: $\De=0.5, Q=0.9,\th=\pi/2$. 
In both cases the early time behavior is well approximated by (\ref{eq:jzExpansion}), indicated by  dashed lines.
}
\label{Fig-Marko}
\end{figure}

\section{Discussion}

We have studied the time evolution of transversal spin helices under coherent $XXZ$ dynamics with arbitrary $z$-axis anisotropy.  
Spin helices are chiral factorized states 
that can be prepared experimentally and they are of potential practical importance. 
We established a fundamental  fact that the helix retains its harmonic space dependence at all times,  which allows to 
describe the density profile as a travelling wave with decaying amplitude. 
We studied the helix profile evolution in the whole phase space via exact  and approximate approaches,  supported
with TEBD.  
 We found that the amplitude decay rate,
 the most important characteristic from a practical viewpoint,  satisfies
scaling relation,  relating decay of homogeneous ($Q= 0$) and a inhomogeneous ($Q\neq 0$) setups.
The scaling relation allows to obtain the  decay rate of a transversal helix with arbitrary wavelength in system with 
arbitrary anisotropy 
from a single scaling  function.  At the free fermion point, the scaling relation holds for all quantum observables.
Further we studied how does the current of magnetization  evolves in time. 
We proposed a criterium for the asymptotic current and pointed out 
open problems of interest for further investigations.  
%From the theoretical viewpoint,  the problem  it is fully describable by a few control parameters and 
%it remains nontrivial also for finite periodic systems.  
We expect that our findings are of direct interest for experimentalists and can serve for calibration purposes in future 
experiments.

\begin{acknowledgments}
VP   acknowledges financial support by the European Research Council through
the advanced grant No. 694544—OMNES, from the Deutsche
Forschungsgemeinschaft through DFG project KL 645/20-2 (VP), and Slovenian Research Agency, Program P1-0402 
and J1-4385 (MZ). XZ acknowledges the financial support from the National Natural Science Foundation of China (No. 12204519).
\end{acknowledgments}

 %\bibliographystyle{apsrev4-2}
%\bibliography{Reference}

\begin{thebibliography}{21}%
\makeatletter
\providecommand \@ifxundefined [1]{%
 \@ifx{#1\undefined}
}%
\providecommand \@ifnum [1]{%
 \ifnum #1\expandafter \@firstoftwo
 \else \expandafter \@secondoftwo
 \fi
}%
\providecommand \@ifx [1]{%
 \ifx #1\expandafter \@firstoftwo
 \else \expandafter \@secondoftwo
 \fi
}%
\providecommand \natexlab [1]{#1}%
\providecommand \enquote  [1]{``#1''}%
\providecommand \bibnamefont  [1]{#1}%
\providecommand \bibfnamefont [1]{#1}%
\providecommand \citenamefont [1]{#1}%
\providecommand \href@noop [0]{\@secondoftwo}%
\providecommand \href [0]{\begingroup \@sanitize@url \@href}%
\providecommand \@href[1]{\@@startlink{#1}\@@href}%
\providecommand \@@href[1]{\endgroup#1\@@endlink}%
\providecommand \@sanitize@url [0]{\catcode `\\12\catcode `\$12\catcode
  `\&12\catcode `\#12\catcode `\^12\catcode `\_12\catcode `\%12\relax}%
\providecommand \@@startlink[1]{}%
\providecommand \@@endlink[0]{}%
\providecommand \url  [0]{\begingroup\@sanitize@url \@url }%
\providecommand \@url [1]{\endgroup\@href {#1}{\urlprefix }}%
\providecommand \urlprefix  [0]{URL }%
\providecommand \Eprint [0]{\href }%
\providecommand \doibase [0]{http://dx.doi.org/}%
\providecommand \selectlanguage [0]{\@gobble}%
\providecommand \bibinfo  [0]{\@secondoftwo}%
\providecommand \bibfield  [0]{\@secondoftwo}%
\providecommand \translation [1]{[#1]}%
\providecommand \BibitemOpen [0]{}%
\providecommand \bibitemStop [0]{}%
\providecommand \bibitemNoStop [0]{.\EOS\space}%
\providecommand \EOS [0]{\spacefactor3000\relax}%
\providecommand \BibitemShut  [1]{\csname bibitem#1\endcsname}%
\let\auto@bib@innerbib\@empty
%</preamble>
\bibitem [{\citenamefont {Hild}\ \emph {et~al.}(2014)\citenamefont {Hild},
  \citenamefont {Fukuhara}, \citenamefont {Schau\ss{}}, \citenamefont {Zeiher},
  \citenamefont {Knap}, \citenamefont {Demler}, \citenamefont {Bloch},\ and\
  \citenamefont {Gross}}]{2014-SHS-Experimental}%
  \BibitemOpen
  \bibfield  {author} {\bibinfo {author} {\bibfnamefont {S.}~\bibnamefont
  {Hild}}, \bibinfo {author} {\bibfnamefont {T.}~\bibnamefont {Fukuhara}},
  \bibinfo {author} {\bibfnamefont {P.}~\bibnamefont {Schau\ss{}}}, \bibinfo
  {author} {\bibfnamefont {J.}~\bibnamefont {Zeiher}}, \bibinfo {author}
  {\bibfnamefont {M.}~\bibnamefont {Knap}}, \bibinfo {author} {\bibfnamefont
  {E.}~\bibnamefont {Demler}}, \bibinfo {author} {\bibfnamefont
  {I.}~\bibnamefont {Bloch}}, \ and\ \bibinfo {author} {\bibfnamefont
  {C.}~\bibnamefont {Gross}},\ }\href {\doibase 10.1103/PhysRevLett.113.147205}
  {\bibfield  {journal} {\bibinfo  {journal} {Phys. Rev. Lett.}\ }\textbf
  {\bibinfo {volume} {113}},\ \bibinfo {pages} {147205} (\bibinfo {year}
  {2014})}\BibitemShut {NoStop}%
\bibitem [{\citenamefont {Jepsen}\ \emph {et~al.}(2022)\citenamefont {Jepsen},
  \citenamefont {Lee}, \citenamefont {Lin}, \citenamefont {Dimitrova},
  \citenamefont {Margalit}, \citenamefont {Ho},\ and\ \citenamefont
  {Ketterle}}]{SHS-Ketterle}%
  \BibitemOpen
  \bibfield  {author} {\bibinfo {author} {\bibfnamefont {P.~N.}\ \bibnamefont
  {Jepsen}}, \bibinfo {author} {\bibfnamefont {Y.~K.}\ \bibnamefont {Lee}},
  \bibinfo {author} {\bibfnamefont {H.}~\bibnamefont {Lin}}, \bibinfo {author}
  {\bibfnamefont {I.}~\bibnamefont {Dimitrova}}, \bibinfo {author}
  {\bibfnamefont {Y.}~\bibnamefont {Margalit}}, \bibinfo {author}
  {\bibfnamefont {W.~W.}\ \bibnamefont {Ho}}, \ and\ \bibinfo {author}
  {\bibfnamefont {W.}~\bibnamefont {Ketterle}},\ }\href
  {https://doi.org/10.1038/s41567-022-01651-7} {\bibfield  {journal} {\bibinfo
  {journal} {Nature Physics}\ }\textbf {\bibinfo {volume} {18}},\ \bibinfo
  {pages} {899} (\bibinfo {year} {2022})}\BibitemShut {NoStop}%
\bibitem [{\citenamefont {Jepsen}\ \emph {et~al.}(2020)\citenamefont {Jepsen},
  \citenamefont {Amato-Grill}, \citenamefont {Dimitrova}, \citenamefont {Ho},
  \citenamefont {Demler},\ and\ \citenamefont
  {Ketterle}}]{2020NatureSpinHelix}%
  \BibitemOpen
  \bibfield  {author} {\bibinfo {author} {\bibfnamefont {P.~N.}\ \bibnamefont
  {Jepsen}}, \bibinfo {author} {\bibfnamefont {J.}~\bibnamefont {Amato-Grill}},
  \bibinfo {author} {\bibfnamefont {I.}~\bibnamefont {Dimitrova}}, \bibinfo
  {author} {\bibfnamefont {W.~W.}\ \bibnamefont {Ho}}, \bibinfo {author}
  {\bibfnamefont {E.}~\bibnamefont {Demler}}, \ and\ \bibinfo {author}
  {\bibfnamefont {W.}~\bibnamefont {Ketterle}},\ }\href {\doibase
  10.1038/s41586-020-3033-y} {\bibfield  {journal} {\bibinfo  {journal}
  {NATURE}\ }\textbf {\bibinfo {volume} {588}},\ \bibinfo {pages} {403+}
  (\bibinfo {year} {2020})}\BibitemShut {NoStop}%
\bibitem [{\citenamefont {Jepsen}\ \emph {et~al.}(2021)\citenamefont {Jepsen},
  \citenamefont {Ho}, \citenamefont {Amato-Grill}, \citenamefont {Dimitrova},
  \citenamefont {Demler},\ and\ \citenamefont
  {Ketterle}}]{2021KetterleTransverse}%
  \BibitemOpen
  \bibfield  {author} {\bibinfo {author} {\bibfnamefont {P.~N.}\ \bibnamefont
  {Jepsen}}, \bibinfo {author} {\bibfnamefont {W.~W.}\ \bibnamefont {Ho}},
  \bibinfo {author} {\bibfnamefont {J.}~\bibnamefont {Amato-Grill}}, \bibinfo
  {author} {\bibfnamefont {I.}~\bibnamefont {Dimitrova}}, \bibinfo {author}
  {\bibfnamefont {E.}~\bibnamefont {Demler}}, \ and\ \bibinfo {author}
  {\bibfnamefont {W.}~\bibnamefont {Ketterle}},\ }\href {\doibase
  10.1103/PhysRevX.11.041054} {\bibfield  {journal} {\bibinfo  {journal} {Phys.
  Rev. X}\ }\textbf {\bibinfo {volume} {11}},\ \bibinfo {pages} {041054}
  (\bibinfo {year} {2021})}\BibitemShut {NoStop}%
\bibitem [{\citenamefont {Rodriguez-Nieva}\ \emph {et~al.}(2022)\citenamefont
  {Rodriguez-Nieva}, \citenamefont {Orioli},\ and\ \citenamefont
  {Marino}}]{2022-FarFromEquilibriumEniversality-2D}%
  \BibitemOpen
  \bibfield  {author} {\bibinfo {author} {\bibfnamefont {J.~F.}\ \bibnamefont
  {Rodriguez-Nieva}}, \bibinfo {author} {\bibfnamefont {A.~P.}\ \bibnamefont
  {Orioli}}, \ and\ \bibinfo {author} {\bibfnamefont {J.}~\bibnamefont
  {Marino}},\ }\href {\doibase 10.1073/pnas.2122599119} {\bibfield  {journal}
  {\bibinfo  {journal} {PROCEEDINGS OF THE NATIONAL ACADEMY OF SCIENCES OF THE
  UNITED STATES OF AMERICA}\ }\textbf {\bibinfo {volume} {119}} (\bibinfo
  {year} {2022}),\ 10.1073/pnas.2122599119}\BibitemShut {NoStop}%
\bibitem [{\citenamefont {Psaroudaki}\ and\ \citenamefont
  {Panagopoulos}(2021)}]{2021SkyrmionsQubits}%
  \BibitemOpen
  \bibfield  {author} {\bibinfo {author} {\bibfnamefont {C.}~\bibnamefont
  {Psaroudaki}}\ and\ \bibinfo {author} {\bibfnamefont {C.}~\bibnamefont
  {Panagopoulos}},\ }\href {\doibase 10.1103/PhysRevLett.127.067201} {\bibfield
   {journal} {\bibinfo  {journal} {Phys. Rev. Lett.}\ }\textbf {\bibinfo
  {volume} {127}},\ \bibinfo {pages} {067201} (\bibinfo {year}
  {2021})}\BibitemShut {NoStop}%
\bibitem [{\citenamefont {K\"uhn}\ \emph {et~al.}()\citenamefont {K\"uhn},
  \citenamefont {Gerken}, \citenamefont {Lena~Funcke}, \citenamefont
  {Stornati}, \citenamefont {Jansen},\ and\ \citenamefont
  {Posske}}]{2023Posske}%
  \BibitemOpen
  \bibfield  {author} {\bibinfo {author} {\bibfnamefont {S.}~\bibnamefont
  {K\"uhn}}, \bibinfo {author} {\bibfnamefont {F.}~\bibnamefont {Gerken}},
  \bibinfo {author} {\bibfnamefont {T.~H.}\ \bibnamefont {Lena~Funcke}},
  \bibinfo {author} {\bibfnamefont {P.}~\bibnamefont {Stornati}}, \bibinfo
  {author} {\bibfnamefont {K.}~\bibnamefont {Jansen}}, \ and\ \bibinfo {author}
  {\bibfnamefont {T.}~\bibnamefont {Posske}},\ }\href
  {https://doi.org/10.48550/arXiv.2302.02603} {\bibinfo  {journal}
  {arXiv:2302.02603}\ }\BibitemShut {NoStop}%
\bibitem [{\citenamefont {Ma}\ \emph {et~al.}(2022)\citenamefont {Ma},
  \citenamefont {Zhang},\ and\ \citenamefont {Song}}]{2022-DM-SHS}%
  \BibitemOpen
\bibfield  {journal} {  }\bibfield  {author} {\bibinfo {author} {\bibfnamefont
  {E.~S.}\ \bibnamefont {Ma}}, \bibinfo {author} {\bibfnamefont {K.~L.}\
  \bibnamefont {Zhang}}, \ and\ \bibinfo {author} {\bibfnamefont
  {Z.}~\bibnamefont {Song}},\ }\href {\doibase 10.1103/PhysRevB.106.245122}
  {\bibfield  {journal} {\bibinfo  {journal} {Phys. Rev. B}\ }\textbf {\bibinfo
  {volume} {106}},\ \bibinfo {pages} {245122} (\bibinfo {year}
  {2022})}\BibitemShut {NoStop}%
\bibitem [{\citenamefont {Posske}\ and\ \citenamefont
  {Thorwart}(2019)}]{2019-PosskeWinding}%
  \BibitemOpen
  \bibfield  {author} {\bibinfo {author} {\bibfnamefont {T.}~\bibnamefont
  {Posske}}\ and\ \bibinfo {author} {\bibfnamefont {M.}~\bibnamefont
  {Thorwart}},\ }\href {\doibase 10.1103/PhysRevLett.122.097204} {\bibfield
  {journal} {\bibinfo  {journal} {Phys. Rev. Lett.}\ }\textbf {\bibinfo
  {volume} {122}},\ \bibinfo {pages} {097204} (\bibinfo {year}
  {2019})}\BibitemShut {NoStop}%
\bibitem [{\citenamefont {Popkov}\ and\ \citenamefont
  {Presilla}(2016)}]{2016SHS-DissCarlo}%
  \BibitemOpen
  \bibfield  {author} {\bibinfo {author} {\bibfnamefont {V.}~\bibnamefont
  {Popkov}}\ and\ \bibinfo {author} {\bibfnamefont {C.}~\bibnamefont
  {Presilla}},\ }\href {\doibase 10.1103/PhysRevA.93.022111} {\bibfield
  {journal} {\bibinfo  {journal} {Phys. Rev. A}\ }\textbf {\bibinfo {volume}
  {93}},\ \bibinfo {pages} {022111} (\bibinfo {year} {2016})}\BibitemShut
  {NoStop}%
\bibitem [{\citenamefont {Popkov}\ and\ \citenamefont
  {Schuetz}(2017)}]{2017SHS-DissGunter}%
  \BibitemOpen
  \bibfield  {author} {\bibinfo {author} {\bibfnamefont {V.}~\bibnamefont
  {Popkov}}\ and\ \bibinfo {author} {\bibfnamefont {G.~M.}\ \bibnamefont
  {Schuetz}},\ }\href {\doibase {10.1103/PhysRevE.95.042128}} {\bibfield
  {journal} {\bibinfo  {journal} {{Phys. Rev. E}}\ }\textbf {\bibinfo {volume}
  {{95}}} (\bibinfo {year} {{2017}}),\
  {10.1103/PhysRevE.95.042128}}\BibitemShut {NoStop}%
\bibitem [{\citenamefont {Popkov}\ \emph {et~al.}(2021)\citenamefont {Popkov},
  \citenamefont {Zhang},\ and\ \citenamefont {Kl\"umper}}]{SHS-Phantom}%
  \BibitemOpen
  \bibfield  {author} {\bibinfo {author} {\bibfnamefont {V.}~\bibnamefont
  {Popkov}}, \bibinfo {author} {\bibfnamefont {X.}~\bibnamefont {Zhang}}, \
  and\ \bibinfo {author} {\bibfnamefont {A.}~\bibnamefont {Kl\"umper}},\ }\href
  {https://link.aps.org/doi/10.1103/PhysRevB.104.L081410} {\bibfield  {journal}
  {\bibinfo  {journal} {Phys. Rev. B}\ }\textbf {\bibinfo {volume} {104}},\
  \bibinfo {pages} {L081410} (\bibinfo {year} {2021})}\BibitemShut {NoStop}%
\bibitem [{\citenamefont {Pereira}\ and\ \citenamefont
  {Mueller}(2022)}]{2022XZ-SHS}%
  \BibitemOpen
  \bibfield  {author} {\bibinfo {author} {\bibfnamefont {D.}~\bibnamefont
  {Pereira}}\ and\ \bibinfo {author} {\bibfnamefont {E.~J.}\ \bibnamefont
  {Mueller}},\ }\href {\doibase 10.1103/PhysRevA.106.043306} {\bibfield
  {journal} {\bibinfo  {journal} {Phys. Rev. A}\ }\textbf {\bibinfo {volume}
  {106}},\ \bibinfo {pages} {043306} (\bibinfo {year} {2022})}\BibitemShut
  {NoStop}%
\bibitem [{\citenamefont {Cecile}\ \emph {et~al.}()\citenamefont {Cecile},
  \citenamefont {Gopalakrishnan}, \citenamefont {Vasseur},\ and\ \citenamefont
  {De~Nardis}}]{SHS-Hydro}%
  \BibitemOpen
  \bibfield  {author} {\bibinfo {author} {\bibfnamefont {G.}~\bibnamefont
  {Cecile}}, \bibinfo {author} {\bibfnamefont {S.}~\bibnamefont
  {Gopalakrishnan}}, \bibinfo {author} {\bibfnamefont {R.}~\bibnamefont
  {Vasseur}}, \ and\ \bibinfo {author} {\bibfnamefont {J.}~\bibnamefont
  {De~Nardis}},\ }\href {https://arxiv.org/abs/2211.03725} {\bibinfo  {journal}
  {arXiv:2211.03725}\ }\BibitemShut {NoStop}%
\bibitem [{\citenamefont {Moudgalya}\ \emph {et~al.}(2022)\citenamefont
  {Moudgalya}, \citenamefont {Bernevig},\ and\ \citenamefont
  {Regnault}}]{2022-QuantumScars-Review}%
  \BibitemOpen
\bibfield  {journal} {  }\bibfield  {author} {\bibinfo {author} {\bibfnamefont
  {S.}~\bibnamefont {Moudgalya}}, \bibinfo {author} {\bibfnamefont {B.~A.}\
  \bibnamefont {Bernevig}}, \ and\ \bibinfo {author} {\bibfnamefont
  {N.}~\bibnamefont {Regnault}},\ }\href {\doibase 10.1088/1361-6633/ac73a0}
  {\bibfield  {journal} {\bibinfo  {journal} {Reports on Progress in Physics}\
  }\textbf {\bibinfo {volume} {85}},\ \bibinfo {pages} {086501} (\bibinfo
  {year} {2022})}\BibitemShut {NoStop}%
\bibitem [{\citenamefont {Zhang}\ \emph {et~al.}(2021)\citenamefont {Zhang},
  \citenamefont {Kl{\"u}mper},\ and\ \citenamefont {Popkov}}]{Phantom-Long}%
  \BibitemOpen
  \bibfield  {author} {\bibinfo {author} {\bibfnamefont {X.}~\bibnamefont
  {Zhang}}, \bibinfo {author} {\bibfnamefont {A.}~\bibnamefont {Kl{\"u}mper}},
  \ and\ \bibinfo {author} {\bibfnamefont {V.}~\bibnamefont {Popkov}},\ }\href
  {https://link.aps.org/doi/10.1103/PhysRevB.103.115435} {\bibfield  {journal}
  {\bibinfo  {journal} {Phys. Rev. B}\ }\textbf {\bibinfo {volume} {103}},\
  \bibinfo {pages} {115435} (\bibinfo {year} {2021})}\BibitemShut {NoStop}%
\bibitem [{\citenamefont {Dzyaloshinskii}(1964)}]{DM1}%
  \BibitemOpen
  \bibfield  {author} {\bibinfo {author} {\bibfnamefont {I.~E.}\ \bibnamefont
  {Dzyaloshinskii}},\ }\href@noop {} {\bibfield  {journal} {\bibinfo  {journal}
  {Sov. Phys. JETP}\ }\textbf {\bibinfo {volume} {19}},\ \bibinfo {pages} {960}
  (\bibinfo {year} {1964})}\BibitemShut {NoStop}%
\bibitem [{\citenamefont {Moriya}(1960)}]{DM2}%
  \BibitemOpen
  \bibfield  {author} {\bibinfo {author} {\bibfnamefont {T.}~\bibnamefont
  {Moriya}},\ }\href {\doibase 10.1103/PhysRev.120.91} {\bibfield  {journal}
  {\bibinfo  {journal} {Phys. Rev.}\ }\textbf {\bibinfo {volume} {120}},\
  \bibinfo {pages} {91} (\bibinfo {year} {1960})}\BibitemShut {NoStop}%
\bibitem [{\citenamefont {Durrett}(2019)}]{durrett2019probability}%
  \BibitemOpen
  \bibfield  {author} {\bibinfo {author} {\bibfnamefont {R.}~\bibnamefont
  {Durrett}},\ }\href@noop {} {\emph {\bibinfo {title} {Probability: theory and
  examples}}},\ Vol.~\bibinfo {volume} {49}\ (\bibinfo  {publisher} {Cambridge
  university press},\ \bibinfo {year} {2019})\BibitemShut {NoStop}%
\bibitem [{\citenamefont {Popkov}\ \emph {et~al.}(2023)\citenamefont {Popkov},
  \citenamefont {Zhang},\ and\ \citenamefont {Kl\"umper}}]{2023ChiralBasis}%
  \BibitemOpen
  \bibfield  {author} {\bibinfo {author} {\bibfnamefont {V.}~\bibnamefont
  {Popkov}}, \bibinfo {author} {\bibfnamefont {X.}~\bibnamefont {Zhang}}, \
  and\ \bibinfo {author} {\bibfnamefont {A.}~\bibnamefont {Kl\"umper}},\
  }\href@noop {} {\bibfield  {journal} {\bibinfo  {journal} {arXiv:2303.14056}\
  } (\bibinfo {year} {2023})}\BibitemShut {NoStop}%
\bibitem [{\citenamefont {Schollw{\" {o}}ck}(2011)}]{Ulrich}%
  \BibitemOpen
  \bibfield  {author} {\bibinfo {author} {\bibfnamefont {U.}~\bibnamefont
  {Schollw{\" {o}}ck}},\ }\href {\doibase
  https://doi.org/10.1016/j.aop.2010.09.012} {\bibfield  {journal} {\bibinfo
  {journal} {Annals of Physics}\ }\textbf {\bibinfo {volume} {326}},\ \bibinfo
  {pages} {96} (\bibinfo {year} {2011})}\BibitemShut {NoStop}%
\end{thebibliography}

%apsrev4-2.bst 2015-08-30 from 4.21a (PWD, AO, DPC/HNN) hacked

\appendix

\section{SHS decay: time asymptotic form in the thermodynamic limit $N\rightarrow \infty$}
\label{App:SHSdecay}

Under usual generic assumptions (the eigenstate thermalization hypothesis),  a $U(1)$- invariant operator like $H$  on an infinite lattice,  acting on a
generic state,  imposes  the $U(1)$ symmetry on any  finite subsystem  asymptotically in time, making the asymptotic 
reduced density matrix of the subsystem
commute with the generator of $U(1)$,  namely
%\begin{align}
%&\lim_{t\rightarrow \infty} \lim_{N\rightarrow \infty} \left[ e^{-i H t} \ket{\Psi} \bra{\Psi} e^{i H t}, U_z \right]=0,\quad U_z =\prod_n \si_n^z
%\label{eq:commU(1)}
%\end{align}
\begin{align}
&\lim_{t\rightarrow \infty} \lim_{N\rightarrow \infty} [\rho_{n_1 n_2 \ldots n_M}(t), \si_{n_1}^z \ldots \si_{n_M}^z]=0, \quad \forall M,  \{n_k\} 
\label{eq:commU(1)}
\end{align}
(note that the limits $t\rightarrow \infty$ and $N\rightarrow \infty$ do not commute),
where $\si_{n_1}^z \ldots \si_{n_M}^z$ is the generator of the $U(1)$ symmetry $U_z =\bigotimes_n \si_n^z$ restricted to  the subsystem. In terms of averages (\ref{eq:observable}),  the Eq.(\ref{eq:commU(1)}) yields
\begin{align}
&\lim_{t\rightarrow \infty} \lim_{N\rightarrow \infty} \langle \sigma _{n_1}^{\alpha_1}\sigma_{n_2}^{\alpha_2}\cdots \sigma_{n_k}^{\alpha_k}(t) \rangle 
=0,\quad \mbox{if $m_{+} \neq m_{-}$},
\label{eq:commU(1)-operators}
\end{align}
where $\al_j=+,-,z$ and $m_{+}$ ($m_{-}$) is the total number of pluses$+$ (minuses $-$) in the upper row of indices in (\ref{eq:commU(1)-operators}).
Indeed only the operators 
$\sigma _{n_1}^{\alpha_1}\sigma_{n_2}^{\alpha_2}\cdots \sigma_{n_k}^{\alpha_k}$ 
with $m_{+} =m_{-}$ commute with $U_z$.  For one and two -point correlations  we have 
\begin{align}
&\lim_{t\rightarrow \infty} \lim_{N\rightarrow \infty} \langle \sigma^{\pm}_{n}(t)\rangle=0, \quad \forall n \label{eq:corr1App}\\ 
&\lim_{t\rightarrow \infty} \lim_{N\rightarrow \infty} \left\{
\langle \sigma_{n}^{\pm}\sigma^{z}_{m}(t)\rangle, \   \langle \sigma_{n}^{+}\sigma^{+}_{m}(t)\rangle, \   \langle \sigma_{n}^{-}\sigma^{-}_{m}(t)\rangle\right\}=0, \quad \forall n,m \label{eq:corr2}
\end{align}
Consequently,
the reduced density matrices for one and two sites written in the computational basis,  asymptotically in time become 
block-diagonal, 
\begin{align}
&\lim_{t\rightarrow \infty} \lim_{N\rightarrow \infty} \rho_{n} (t)= \frac12 \left(
\begin{array}{cc}
a &0 \\
0 &d
\end{array}\right),
 \quad \forall n \label{eq:ro1}\\ 
&\lim_{t\rightarrow \infty} \lim_{N\rightarrow \infty} \rho_{n,m}(t)= \left(
\begin{array}{cccc}
a &0 &0 &0\\
0 &b &b_1 &0\\
0 &b_1^* &c &0\\
0 &0 &0 &d
\end{array}\right),
 \quad \forall n,m. \label{eq:ro2}
\end{align}
%where $j^z(\infty) = 4i (b_1 - b_1^*)$,   is a $U(1)$-invariant quantity. 
Generic form of asymptotic in time reduced density matrix for arbitrary sites $M$ is blockdiagonal as in (\ref{eq:ro2}),  and satisfies (\ref{eq:commU(1)}).

\section{Symmetries of SHS amplitude and  phase  $S_N(Q,\th,\De,t)$ and $\phi(Q,\th,\De,t)$ }
\label{App:PhaseShift}

Consider operator $U_x= \bigotimes_{n=-\infty}^\infty \si_n^x$, and the operator $R$ of mirror reflection wrt. the middle site $0$ which has the property
\begin{align}
&U_x \ket{\Psi_{Q,\th}}= \ket{\Psi_{-Q,\pi-\th}},\\
&R \ket{\Psi_{Q,\th}} = \ket{\Psi_{-Q,\th}},
\end{align}
where $\th$ is the polar angle characterizing the SHS. 
Using the properties $[H,U_x]=0$,
$R^\dagger = R$,  $U_x^\dagger = U_x$, we obtain
\begin{align}
&\langle{A}\rangle_{Q,\pi -\th}= \langle U_x A U_x\rangle_{-Q,\th},\\
&\langle{A}\rangle_{-Q}= \langle RA R\rangle_Q .
\end{align}
Here and below we shall use the shorthand notations $ \langle A \rangle_{Q,\th}$, $\langle A \rangle_{Q,\vfi}$, $\langle A \rangle_{Q}$ 
to denote 
generic expectation of an operator $A$:  
$\langle A(H,t)\rangle_{Q,\th,\vfi}  =\bra{\Psi_{Q,\th,\vfi}} e^{i H t}  \  A \  e^{-i H t} \ket{\Psi_{Q,\th,\vfi}}$
%{$\langle A(H,t)\rangle_{Q,\th,\vfi}$ in Eq. (\ref{eq:observable})}  
(e.g. using $\langle A \rangle_{Q}$ means that $\th$ and $\vfi$ are the same on both sides of an equality).

In particular, for one-point correlations we obtain
 \begin{align}
&\langle\si_n^x\rangle_{Q,\pi-\th}= \langle \si_n^{x} \rangle_{-Q,\th}, \label{eq:UxActionX}\\
&\langle{\si_n^{y,z}}\rangle_{Q,\pi-\th}= -\langle {\si_n^{y,z}}\rangle_{-Q,\th},  \label{eq:UxActionXY}\\
&\langle{\si_0^{\al}}\rangle_{-Q}=\langle {\si_0^{\al}}\rangle_{Q}, \quad \al=x,y,z \label{eq:RActionSite0},
\end{align}
the last one following from $R \,\si_0^{\al} R=\si_0^{\al}$, since mirror reflection $R$ does not touch the central site.  

Comparing an identity $\langle{\si_n^+}\rangle_{\pi-\th} = \langle{R \,\si_n^- R}\rangle_{\th}=\langle{ \si_{-n}^- }\rangle_{\th}$ and comparing to (\ref{eq:SigmaX}), (\ref{eq:SigmaY}), we obtain 
\begin{align}
S_N(\pi-\th) = S_N(\th),\quad \phi(\pi-\th) = -\phi(\th).
\end{align}
Analogously, analyzing the identity $\langle{\si_n^+}\rangle_Q= \langle{R \,\si_n^+ R}\rangle_{-Q}=\langle{ \si_{-n}^+ }\rangle_{-Q} $
\begin{align}
S_N(Q) = S_N(-Q),\quad &\phi(Q) = \phi(-Q).\label{Symmetry:Q}
\end{align}
Since $\langle \sigma_n^x(t)\rangle_{Q,\vfi}$ is a real number,  we obtain
\begin{align}
\langle \sigma_n^x(t)\rangle_{Q,\vfi}&=\bra{\Psi_{Q,\vfi}} e^{iHt}  \sigma_n^x \,e^{-iHt}  \ket{\Psi_{Q,\vfi}}\no\\
&=\bra{\Psi_{-Q,-\vfi}} e^{-iHt}  \sigma_n^x \,e^{iHt}  \ket{\Psi_{-Q,-\vfi}}\no\\
&=\langle \sigma_n^x(-t)\rangle_{-Q,-\vfi}.
\end{align}
With the help of Eq. (\ref{Symmetry:Q}), one can prove
\begin{align}
&S_N(Q,t) = S_N(-Q,t)=S_N(Q,-t),\label{Symmetry:t1}\\&\phi(Q,t) = \phi(-Q,t)=-\phi(Q,-t).\label{Symmetry:t2}
\end{align}
For an even $N$,  if $Q$ satisfies (\ref{eq:commensurate}), $\pi\pm Q$ satisfy  (\ref{eq:commensurate}) as well. 
Using  (\ref{eq:Theorem}), one can prove that when $Q\to Q+\pi$
\begin{align*}
&H'|_{Q\to Q+\pi}=-H'|_{\De\to -\De},\\
&(\sigma_n^{\pm})'|_{Q\to Q+\pi}=e^{\pm i\pi n}(\sigma_n^{\pm})',\\
&\langle \sigma_n^+(H,t)\rangle_{Q+\pi}=e^{\pm i\pi n}\langle \sigma_n^+(H|_{\De\to -\De},-t)\rangle_Q,
\end{align*}
which leads to the  relations
\begin{align}
&S_N(Q, \De, t) = S_N(-Q,\De,t)= S_N(\pi-Q, -\De, -t)=S_N(\pi-Q, -\De, t),\\
&\phi(Q,\De, t) = \phi(-Q, \De, t)=\phi(\pi-Q, -\De,-t)=-\phi(\pi-Q,-\De, t).
\end{align}
Selected relations of this Appendix are quoted in the text,  omitting the repeated symbols for brevity.

\section{Proof of Eq. (\ref{eq:PhaseVelocity})}
\label{App:v(0)}

Consider a time evolution of a density matrix  in the form of a homogeneous diagonal factorized state
perturbed by a nondiagonal term with wavevector $Q$:
\begin{align}
&\rho_\eps(0)=R^{\otimes_N} + \eps \sum_n A_n  \label{B0} \\ 
&A_n = R^{\otimes_{n-1}} \otimes F_n \otimes  R^{\otimes_{N-n}} \nonumber \\
& R = \left(
\begin{array}{cc}
a &0  \\
0 &d 
\end{array}\right), \quad a = \frac12 + \frac{\cos \th}{2},\ d=1-a \nonumber\\
& F_n = \left(
\begin{array}{cc}
0 &e^{-iQn}  \\
 e^{iQn} &0 
\end{array}\right), \nonumber
\end{align}

The time evolved state is 
\begin{align}
&\rho_\eps(t) = \rho_\eps(0) -i t  [H, \rho_\eps(0)] - \frac{t^2}{2} ad_H^2 \rho_\eps(0) + \ldots \nonumber \\
&=  \rho_\eps(0) -i \eps t \sum_n [h_{n,n+1}, A_n +A_{n+1}] + O(t^2)= \nonumber\\
& =    \rho_\eps(0) -i \eps t \sum_n  R^{\otimes_{n-1}} \otimes [h, F_n\otimes R  +R \otimes F_{n+1}]\otimes  R^{\otimes_{N-n-1}} + O(t^2),\label{B1}
\end{align}
where $h= \si^x \otimes \si^x + \si^y \otimes \si^y+ \De (\si^z \otimes \si^z-I) $ is the energy density of the Hamiltonian $H$
and we used $[h,R\otimes R]=0$.
Denoting 
\begin{align}
&X'= \si^z X,
\end{align}
and substituting easily verifyable relations
\begin{align}
&[h,F\otimes R] = 2 \De (F' \otimes R') - 2(R' \otimes F') \nonumber\\
&[h,R\otimes F] = 2 \De (R' \otimes F') - 2(F' \otimes R')\nonumber
\end{align}
into (\ref{B1}), we readily obtain
\begin{align}
&\rho_\eps(t) =\rho_\eps(0) -i \eps t \sum_n  Z_{n,n+1}+ O(t^2), \label{B2}\\
&Z_{n,n+1} = 
2 R^{\otimes_{n-1}} \otimes 
\left(
\De F_n' \otimes R' - R' \otimes F_{n}' + \De  R' \otimes F_{n+1}' -  F_{n+1}' \otimes R'
\right)
\otimes  R^{\otimes_{N-n-1}}.\nonumber
\end{align}
Let us now calculate the observable $\langle \si_n^+(t) \rangle$ using  (\ref{B2}).
\begin{align}
&\langle \si_n^+(t) \rangle = \langle \si_n^+(0) \rangle - i \eps t \sum_{m=n-1}^n tr( Z_{m,m+1} \ \si_n^+)+ O(t^2).\
\label{B3}
\end{align}
Substituting 
\begin{align}
&tr(Z_{n-1,n} \si_n^+ ) = 2 tr ((-F_{n-1}'+ \De F_n')\, \si_n^+) \ tr(RR') = 
2 (e^{iQ (n-1)} - \De e^{iQn}) \cos \th,\\
&tr(Z_{n,n+1} \si_n^+ ) = 
2 (-\De  e^{iQ n} +  e^{iQ(n+1)}) \, \cos \th,
\end{align}
into (\ref{B3}) we finally obtain
\begin{align}
&\langle \si_n^+(t) \rangle = \eps e^{i Q n } \left(
1- 2 i t \left( 
e^{iQ} + e^{-i Q} -2 \De
\right) \cos \th
\right) + O(t^2)\no\\
& \approx \eps e^{iQ n } e^{-4it (\cos Q - \De) \cos \th}\no\\
&= \eps e^{i (Q n - \phi(t))}=  \eps e^{i (Q (n - v_0 t)},\\
&v_0 = \frac{4 (\cos Q - \De) \cos \th}{Q},
\end{align}
where $v_0$ is the phase velocity of the infinitesimal perturbations.   
Quite remarkably,  our rather simple-minded analysis turns out to predict qualitatively correctly 
the initial phase velocity $v_0=\phi'(0)/Q $,  even though the $Q$-dependent amplitude is not infinitesimally small,  
  see Fig. \ref{Fig-PhaseVelocity} of the main text.

\section{$S_N(t)$ and $\phi(t)$ for $\theta\to 0$ case}\label{Theta0}
%Here, $a=\tan\frac{\theta}{2}$ and $a\to0,\infty$ while $\theta\to 0,\pi$. 
$\ket{\Psi_{Q,\theta,\vfi}}$ is a linear combination of $N+1$ independent states: $\ket{\xi_0},\,\ket{\xi_1},\ldots,\ket{\xi_N}$ \cite{SHS-Phantom}
\begin{align}
&\ket{\Psi_{Q,\theta,\vfi}}=\eE^{-\frac{i}{2}(N\vfi+Q\sum_{n=0}^{N-1} n)}\cos^N\!\tfrac{\theta}{2}\,\sum_{m=0}^N\tan^m\!\tfrac{\theta}{2}\,\eE^{im\vfi}\ket{\xi_m},\label{Phantom;1}\\ &\ket{\xi_m}=\sum_{k_1<k_2<\ldots<k_m}\eE^{iQ\sum_{j=1}^m k_j}\sigma_{k_1}^-\cdots\sigma_{k_m}^-\,\ket{\uparrow\uparrow\cdots\uparrow}.
\end{align}
In the $\theta\to0$ limit, $\tan\frac{\theta}{2}\to 0$, therefore we only consider the first two leading terms, i.e. $\ket{\xi_0}$ and $\ket{\xi_1}$. Since $\eE^{iNQ}=1$, one can prove that both $\ket{\xi_0}$ and $\ket{\xi_1}$ are eigenstates of $H$
\begin{align}
H\ket{\xi_1}=4(\cos Q-\De)\ket{\xi_1},\quad H\ket{\xi_0}=0.
\end{align}
Then, we have
%\begin{align}&S^{\al  }_{n}(t) = \bra{SHS_{Q,\th,u}} e^{i H t}  \  \si_{n}^\al  e^{-i H t} \ket{SHS_{Q,\th,u}},\quad \al= x,y \label{eq:observableApp}\end{align}
\begin{align}
\langle \sigma_n^x( t)\rangle&=\bra{\Psi_{Q,\theta,\vfi}}\eE^{iHt}\sigma_n^x\,\eE^{-iHt}\ket{\Psi_{Q,\theta,\vfi}}\no\\
&=\cos^{2N}\!\tfrac{\theta}{2}\left(\bra{\xi_0}+\tan\tfrac{\th}{2}\eE^{-i\vfi+4i(\cos Q-\De)t}\bra{\xi_1}\right)\sigma_n^x\left(\ket{\xi_0}+\tan\tfrac{\th}{2}\eE^{i\vfi-4i(\cos Q-\De)t}\ket{\xi_1}\right)+\cdots\no\\
&=\cos^{2N}\!\tfrac{\theta}{2}\tan\tfrac{\th}{2}\left(\eE^{i\vfi-4i(\cos Q-\De)t}\bra{\xi_0}\sigma_n^x\ket{\xi_1}+\eE^{-i\vfi+4i(\cos Q-\De)t}\bra{\xi_1}\sigma_n^x\ket{\xi_0}\right)+\cdots\no\\
%&=\frac{2a}{(1+a^2)^N}\cos(4t(\cos Q-\De)+(2u-n)Q)+\cdots\no\\
&=2\cos^{2N}\!\tfrac{\theta}{2}\tan\tfrac{\th}{2}\cos\left(nQ-4t(\cos Q-\De)+\vfi\right)+\cdots,\\[2pt]
%%%%%%%%%%%%%%%%%%%%%%%%%%%%%%%%%%%%%%%%%%%%%
\langle \sigma_n^y( t)\rangle&=\bra{\Psi_{Q,\theta,\vfi}}\eE^{iHt}\sigma_n^y\,\eE^{-iHt}\ket{\Psi_{Q,\theta,\vfi}}\no\\
&=\cos^{2N}\!\tfrac{\theta}{2}\left(\bra{\xi_0}+\tan\tfrac{\th}{2}\eE^{-i\vfi+4i(\cos Q-\De)t}\bra{\xi_1}\right)\sigma_n^y\left(\ket{\xi_0}+\tan\tfrac{\th}{2}\eE^{i\vfi-4i(\cos Q-\De)t}\ket{\xi_1}\right)+\cdots\no\\
&=\cos^{2N}\!\tfrac{\theta}{2}\tan\tfrac{\th}{2}\left(\eE^{i\vfi-4i(\cos Q-\De)t}\bra{\xi_0}\sigma_n^y\ket{\xi_1}+\eE^{-i\vfi+4i(\cos Q-\De)t}\bra{\xi_1}\sigma_n^y\ket{\xi_0}\right)+\cdots\no\\
%&=\frac{2a}{(1+a^2)^N}\cos(4t(\cos Q-\De)+(2u-n)Q)+\cdots\no\\
&=2\cos^{2N}\!\tfrac{\theta}{2}\tan\tfrac{\th}{2}\sin\left(nQ-4t(\cos Q-\De)+\vfi\right)+\cdots,
\end{align}
which leads to the following asymptotic behavior  in the limit $\theta\to 0$:
\begin{align}
&\lim_{\theta\to 0}\frac{\langle \sigma_n^x(t)\rangle}{\sin\theta}=\cos\left(nQ-4t(\cos Q-\De)+\vfi\right),\label{Theta0;1}\\
&\lim_{\theta\to 0}\frac{\langle \sigma_n^y(t)\rangle}{\sin\theta}=\sin\left(nQ-4t(\cos Q-\De)+\vfi\right),\label{Theta0;2}\\
&\lim_{\theta\to 0}{S_N(t)}=1,\label{Theta0;3}\\
&\lim_{\theta\to 0}{\phi(t)}=4t(\cos Q-\De).\label{Theta0;4}
\end{align}

\section{Ising case}
\label{App:Ising}

Here we find the SHS amplitude  time dependence,  for $\De \gg 1$ limit.
The rescaled amplitude $S_N$ is given by
\begin{align}
&S_N(t)=\frac{2}{\sin\theta}\sqrt{\langle \si_0^+(t) \rangle \langle \si_0^{-}(t) \rangle},
\label{eq:S(t)app}
\end{align}
For large $\De \gg 1$,  in the zero order approximation,  we neglect the hopping part of the Hamiltonian
\begin{align}
&\langle \si_0^+(t) \rangle = \bra{\Psi_{Q}} e^{i H_{00z} t}  \  \si_{0}^+ \   e^{-i H_{00z} t} \ket{\Psi_{Q}},\label{eq:SigmaPlus}
\end{align}
where $H_{00z}= \De \sum_{n=0}^{N-1} \si_n^z  \si_{n+1}^z$. 
Using $\ket{\Psi_{Q}} = U_Q \ket{\Psi_{0}}$, with $U_Q= e^{-i \frac{Q}{2} \sum_{n=0}^{N-1} n \, \si_n^z}$
we obtain,  using diagonality of  $H_{00z}$,  and $[U,\si_0^\al]=0$:
\begin{align*}
&\bra{\Psi_{Q}} e^{i H_{00z} t}  \  \si_{0}^+ \  e^{-i H_{00z} t} \ket{\Psi_{Q}} \\
&= \bra{\Psi_{0}} U_Q^\dagger e^{i H_{00z} t}  \  \si_{0}^+ \ e^{-i H_{00z} t}  U_Q \ket{\Psi_{0}}\\
&=\bra{\Psi_{0}}  e^{i H_{00z} t}  \  \si_{0}^+ \   e^{-i H_{00z} t}   \ket{\Psi_{0}}\\
&= \bra{\Psi_{0}}  V_{N-1,0}^\dagger  V_{01}^\dagger \si_{0}^+ V_{01}  V_{N-1,0} \ket{\Psi_{0}},\quad 
V_{n,m} = e^{-i \De t \si_n^z \si_m^z}.
%&  \ket{SHS_{0,\th}} = \binom{a}{b}^{\otimes_N}, \quad a=\cos\frac{\th}{2}, \ b =\sin\frac{\th}{2}.
\end{align*}
After straighforward  calculation involving three qubits located at consecutive positions $N-1,0,1$,  we obtain
\begin{align}
&\langle \si_0^+(t) \rangle = \frac{1}{8} e^{i \vfi -4 i \Delta  t} \sin \theta\left[1-\cos\theta+(1+\cos \theta)e^{4 i \Delta  t}\right]^2.
\end{align}
Complex conjugation of the  above gives $\langle \si_0^-(t) \rangle$.  Substituting in 
(\ref{eq:S(t)app}),  after some algebra we obtain
\begin{align}
&S_N(t)=  \frac12 \left[ 1+ \cos^2\th + \sin^2 \th  \cos (4 \De t) \,  \right].
\label{eq:S(t)Ising}
\end{align}
This describes harmonic motion with  a positive nonzero mean. Note that there is no dependence on the wavelength $Q$.
For fixed $\th\neq \pi/2$ (SHS out of $XY$-plane),  the amplitude stays always positive, 
\begin{align}
&\min_t S_N(t)=\cos^2 \th
\label{eq:S(t)IsingMin}
\end{align}
The presence of the small hopping term in the Hamiltonian,  neglected in our calculation,  for large but finite $\De$
leads to (A): slow gradual decrease of the amplitude $S(t)$  in accordance with $U(1)$ symmetry restoration principle
Eq.(\ref{eq:ro1}),
and (B): to
$1/\De$ corrections to (\ref{eq:S(t)Ising}),  which can be incorporated into a finite shift of $\De$  value in 
(\ref{eq:S(t)Ising}).  For $Q=0, \ \th=\pi/2$ we found $\De  \rightarrow \De -1.3$ effective shift by a comparison to numerical data
for large $\De$ (data not shown).

\section{Taylor expansion for $N \rightarrow \infty$}
\label{App:Taylor}

The operator expansion

\begin{align}
&e^X A e^{-X}  =A + [X,A] + \frac{1}{2!} [X,[X,A]] + \ldots = \sum_{n=0}^\infty \frac{1}{n!} ad^n_X(A) \label{eq:operatorExpansion1}
\\
&ad_X(A)=[X,A], \quad ad^0_X(A) = A \label{eq:adX}
\end{align}
with $X$ substituted by $i H t$ and $A$ being the operator of a chosen observable,
allows to find Taylor expansions,  up to certain order,  for any observable in the thermodynamic limit.

From (\ref{eq:operatorExpansion}),
substituting $X \rightarrow i H t$, we obtain 
\begin{align}
&\langle A(t) \rangle = \langle A(0) \rangle + \sum_{k>0} C_k t^k.\label{eq:Taylor1}
\end{align}
To determine the $C_k$ in (\ref{eq:Taylor1}), we embed the operator $A$ close to site $n=0$ and 
calculate the term by using symbolic calculations (Mathematica)
\begin{align}
&C_k =\bra{\Psi_{Q,\th,\vfi}}  \frac{1}{k!} ad^k_{i H}(A) \ket{\Psi_{Q,\th,\vfi}},
\end{align}
where $\ket{\Psi_{Q,\th,\vfi}}$ is defined in Eq. (\ref{eq:SHS}) and $Q$ satisfies (\ref{eq:commensurate}). The system size $N$ must be chosen sufficiently large to  exclude
the finite size effects.  For an operator $A$ embedded on a cluster of sites $n  \in [-f,f]$,  $[H,A]$ is embedded on a cluster  $n  \in [-f-1,f+1]$,
and $ad^k_{i H}(A)$ is embedded on a cluster  $n  \in [-f-k,f+k] = [-n_L,n_R]$.  To guarantee the absence of finite size effects,
the cluster  $[-n_L,n_R]$ must lie entirely  inside the segment $[-\frac{N}{2},\frac{N}{2}]$, i.e. 
\begin{align}
&\frac{N}{2}-1 \geq n_R,  \quad -\frac{N}{2}\leq -n_L  \label{app:Nchoice}
\end{align}
For one-point correlations embedded on site $n=0$,  $n_L=n_R=k$,  and we have $N \geq 2(k+1)$  from (\ref{app:Nchoice}). 

Finally,  for products of observables $\langle A \rangle \langle B \rangle$,  a further resummation of the Taylor expansion must be made. 
The Taylor expansion of the amplitude  $S^2(t) = \frac{4}{\sin^2\th} \langle \si_0^+ \rangle \langle \si_0^- \rangle$ 
for $Q=\De=0$ thus obtained is 
\begin{align}
& S^2(\th, t) =1 + 8 \sin^2 \theta\left( -  t^2 + \frac{ t^4}{3}  (3
   \cos (2 \theta )+17) - \frac{2  t^6  }{45}  (156 \cos (2 \theta )+5 \cos (4 \theta )+511) \right.  \nonumber \\
&\left.  +\frac{t^8}{315}  (6777 \cos (2 \theta )+758 \cos (4 \theta )+7 \cos (6 \theta )+19914)  \right.  \nonumber \\
&\left.  -\frac{ 2 \, t^{10}}{14175} (256792 \cos (2 \theta )+68804 \cos (4 \theta )+2792 \cos (6 \theta )+11
   \cos (8 \theta )+916529) \right. \nonumber \\
&\left. +\frac{2 t^{12}}{467775} (5602870 \cos (2 \theta )+5557480 \cos (4 \theta )+460051 \cos (6 \theta )+11410 \cos (8\theta )+55 \cos (10 \theta )+48757510) \right)+
O(t^{14}),
\label{eq:SSthetaExpansion}
\end{align}
(further terms of the expansions  are too bulky to be reported).  It can be verified that for $\th=\pi/2$ the (\ref{eq:SSthetaExpansion})
is compatible with the  Taylor expansion given in the main text. 

With this method we obtain all the other Taylor expansions in the thermodynamic limit quoted in the main text.
Note that as usual, the Taylor expansion (\ref{eq:Taylor1}) has a finite radius of convergence.

\section{Numerical time-evolution}
\label{App:DMRG}

To evolve the initial SHS state $\ket{\Psi_Q}$ with unitary propagator $U(t)={\rm e}^{-{\rm i} H t}$ we write the state in the matrix-product state (MPS) form.
\begin{equation}
 \ket{\Psi_Q}=\sum_{\mathbf{s}} c_{\mathbf{s}} \ket{s},\qquad c_{\mathbf{s}=s_1,\ldots,s_N}=\langle 1 | M_1^{(s_1)} M_2^{(s_2)}\cdots M_N^{(s_N)} | 1 \rangle,
\end{equation}
in terms of two matrices $M_j^{s}$ of size $\chi \times \chi$ for each site $j$, and use standard TEBD method~\cite{Ulrich}. In brief, we split the nearest-neighbor $H=A+B$ into non-commuting $A$ and $B$, each acting on even and odd pairs of spins, respectively. Time evolution is split into small time steps of length $dt$, for which we then Trotterize the propagator into terms involving only $A$, or only $B$, each involving commuting nearest-neighbor transformations. While higher order Trotter-Suzuki schemes are advantageous, we here use a simple leapfrog scheme, $U(dt) \approx {\rm e}^{-{\rm i} A dt/2} {\rm e}^{-{\rm i} B dt} {\rm e}^{-{\rm i} A dt/2}$. Applying one nearest-neighbor unitary transformation acting on sites $j$ and $j+1$ mixes the MPS form on those two sites, which is then restored after doing a singular value decomposition. Time complexity of simulating one unit of time (that consists of $\sim 1/dt$ leapfrog steps) scales as $\sim \chi^3 n/dt$, while the error due to the leapfrog scheme scales as $\sim (dt)^2$. Keeping the MPS representation exact would require the bond size $\chi$ to be equal to the rank of the reduced density operator. Because that rank quickly grows in $t$, and soon saturates at its maximal value that is exponential in $N$, one needs to truncate matrices to some fixed maximal size $\chi$. How large truncation errors due to this finite $\chi$ are then depends on the spectrum of the reduced density matrix. Roughly speaking on can estimate the required $\chi$ to be exponential in the von Neumann entropy. Because von Neumann entropy will generically grow linearly in time this means that the maximal reliable time up-to which we can simulate unitary evolution scales as $\ln \chi$; even for $\chi \sim 10^3$ this is rather small.

As an example, in Fig.~\ref{Fig-Marko}(a) we have an example where the prefactor in the linear growth of von Neumann entropy is large -- one has $S(t=3) \approx 6$ and $S(t=4)\approx 8$, meaning e.g. that with $\chi=256$ one can simulate up-to $t\approx 3$, with $\chi=1024$ only up-to $t\approx 4$. Increasing $\chi$ from $256$ to $1024$ increases CPU time by $\approx 4^3=64$ while bringing in only one additional unit of time. If entanglement is smaller, an example would be $Q=0.9$ in Fig.~\ref{Fig-Marko}(b), one can go to longer times and get more precise results, however, ultimately one again runs into an ``entanglement'' barrier where no further simulation in time is feasible. We also note that in Fig.~\ref{Fig-Marko}(a) where the current varies a lot we could use Trotter timestep $dt=0.05$, while in Fig.~\ref{Fig-Marko}(b) where the current changes much less on the shown time $t \approx 10$, and we therefore wanted the errors to be much less than $1\%$, we had to use $dt=0.01$.

\end{document}